\documentclass[aps,prl,twocolumn,superscriptaddress]{revtex4-1}
\usepackage{amssymb}
\usepackage{epsfig}
\usepackage{graphicx,amsmath,bm}
\usepackage{color}
\usepackage{amsmath}
\usepackage{amsfonts}
\usepackage{graphicx}

\def\bi#1\ei {\begin{itemize}#1\end{itemize}}
\def\bea#1\eea {\begin{align}#1\end{align}}
\def\bean#1\eean {\begin{align*}#1\end{align*}}
\def\ben#1\een {\begin{equation*}#1\end{equation*}}
\def\be#1\ee {\begin{equation}#1\end{equation}}
\def\bes#1\ees {\begin{equation}\begin{split}#1\end{split}\end{equation}}

\newcommand{\la}{\langle} 
\newcommand{\ra}{\rangle}

\begin{document}
\title{Topology by Dissipation in Atomic Quantum Wires}

\author{S. Diehl} 
\affiliation{Institute for Quantum Optics and Quantum Information of the Austrian Academy
of Sciences, A-6020 Innsbruck, Austria}
\affiliation{Institute for Theoretical Physics, University of Innsbruck, A-6020 Innsbruck, Austria}

\author{E. Rico}
\affiliation{Institute for Quantum Optics and Quantum Information of the Austrian Academy
of Sciences, A-6020 Innsbruck, Austria}
\affiliation{Institute for Theoretical Physics, University of Innsbruck, A-6020 Innsbruck, Austria}

\author{M. A. Baranov}
\affiliation{Institute for Quantum Optics and Quantum Information of the Austrian Academy
of Sciences, A-6020 Innsbruck, Austria}
\affiliation{Institute for Theoretical Physics, University of Innsbruck, A-6020 Innsbruck, Austria}
\affiliation{RRC ``Kurchatov Institute'', Kurchatov Square 1, 123182 Moscow, Russia}

\author{P. Zoller}
\affiliation{Institute for Quantum Optics and Quantum Information of the Austrian Academy
of Sciences, A-6020 Innsbruck, Austria}
\affiliation{Institute for Theoretical Physics, University of Innsbruck, A-6020 Innsbruck, Austria}

\begin{abstract}
Robust edge states and non-Abelian excitations are the trademark of topological states of matter, with promising applications such as ``topologically protected'' quantum memory and computing. While so far topological phases have been exclusively discussed in a Hamiltonian context, we show that such phases and the associated topological protection and phenomena also emerge in open quantum systems with engineered dissipation. The specific system studied here is a quantum wire of spinless atomic fermions in an optical lattice coupled to a bath. The key feature of the dissipative dynamics described by a Lindblad master equation is the existence of \emph{Majorana edge modes}, representing a non-local decoherence free subspace. The isolation of the edge states is enforced by a dissipative gap in the p-wave paired bulk of the wire. We describe dissipative non-Abelian braiding operations within the Majorana subspace, and we illustrate the insensitivity to imperfections. Topological protection is granted by a nontrivial winding number of the system density matrix.
\end{abstract}

\maketitle

Topological properties can protect quantum systems from microscopic details and imperfections. In condensed matter physics this is illustrated by the seminal examples of the quantum Hall effect and the recently discovered topological insulators \cite{KaneMele,Bernevig,Zhang,Nayak08,FuKane,HasanKane}. 
The ground state of the Hamiltonian of such systems is characterized by nonzero values of topological invariants which imply the existence of robust  edge states in interfaces to topologically trivial phases. Due to their topological origin, these modes are immune against a wide class of perturbations.

The conceptually simplest example illustrating these phenomena is Kitaev's quantum wire representing a topological superconducting state supporting Majorana fermions as edge states \cite{Kitaev00}. The pair of Majorana edge modes represents a nonlocal fermion which is a promising building block to encode topological qubits \cite{Sau,Beenakker,Alicea11}. Similar to the Majorana excitations near vortices of a $p_{x}+ip_{y}$ superconductor \cite{ReadGreen,Ivanov}, they show nonabelian exchange statistics when braided in 1D wire networks \cite{Alicea11}.

Remarkably, the above described topological features and phenomena not only occur as properties of  Hamiltonians, but appear also in driven dissipative quantum systems. Below we will develop such a topological program for a dissipative many-body system parallel to the Hamiltonian case. We will do this for a dissipative version of Kitaev's quantum wire. This represents the simplest instance exhibiting the key features such as dissipation induced Majorana edge modes, decoupled from the dynamically created p-wave superfluid bulk by a dissipative gap. The dissipation induced topological order is generated in stationary states far away from thermodynamic equilibrium which are not necessarily pure, i.e. described in terms of a wave function only, and is reached exponentially fast from arbitrary initial states. This is in marked contrast to recent ideas of topological order in Hamiltonian systems under non-equilibrium periodic driving conditions \cite{Lindner11,Kitagawa10}. Such a system can be realized with cold atoms in optical lattices, where the generation of topological order in the more conventional Hamiltonian settings has been proposed recently in a variety of settings \cite{Zhang07,Stanescu09,Goldman10,Bermudez10,Jiang10}. 


\section{Dissipative edge modes in a fermionic quantum wire}

Our goal is to develop a master equation for a dissipative quantum wire which exhibits topological properties including Majorana edge states. To illustrate the analogies and differences to the Hamiltonian case, and in particular to motivate our construction of the master equation with the topological states as  dark steady states, we start by briefly summarizing Kitaev's model of the topological superconductor. 

\emph{Topological quantum wire} -- Kitaev considers spinless fermions $a_{i},a_{i}^{\dag }$ on a finite chain of $N$ sites $i$ described by a Hamiltonian 
\begin{equation*}
H=\sum_{i=1}^{N}\left[ -Ja_{i}^{\dag }a_{i+1}+(\Delta a_{i}a_{i+1} + \text{h.c.}) -\mu a_{i}^{\dag }a_{i}  \right],
\end{equation*}
with a hopping term with amplitude $J$, a pairing term with order parameter $\Delta $, and a chemical potential $\mu $. The topologically non-trivial phase of the model is best illustrated for the choice of parameters $J=|\Delta|$ and $\mu =0$, where the Hamiltonian simplifies to 
\begin{equation}
H=2\,\mathrm{i}J\sum_{i=1}^{N-1}c_{2i}\,c_{2i+1}=2J\sum_{i=1}^{N-1}\tilde{a}_{i}^{\dag }\tilde{a}_{i}.  \label{HKit}
\end{equation}
Here we have defined Majorana operators $c_{i}$ as the quadrature components of the complex fermion operators $a_{i}=\frac{1}{2}\left(\mathrm i c_{2i-1}+c_{2i}\right) $ with properties $c_{j}^{\dag}=c_{j},\{c_{j},c_{l}\}=2\delta _{jl}$. The Hamiltonian is readily diagonalized in terms of fermionic Bogoliubov quasiparticle operators $\tilde{a}_{i}=\frac{1}{2}\left( c_{2i}+\mathrm i c_{2i+1}\right) $, where importantly the pairing of Majoranas is from \emph{different physical sites}.

The ground state satisfies the condition $\tilde{a}_{i}|G\rangle =0$ for all $i$. The bulk of the wire describes a fermionic $p$-wave superfluid with a \emph{bulk spectral gap}, which here equals the constant dispersion $\epsilon _{k}=2J$. For a finite wire, the absence of the term $\tilde{a}_{N}^{\dag }\tilde{a}_{N}=\mathrm{i}c_{2N}c_{1}$ indicates the existence of a two-dimensional \emph{zero energy non-local} fermionic subspace spanned by $|\alpha \rangle \in \{ |0\rangle ,|1\rangle =\tilde{a}_{N}^{\dag }|0\rangle\} $. While highly delocalized in terms of the original complex fermion, in the real Majorana basis the situation is described in terms of two Majorana edge modes $\gamma_{L}=c_{1}$ ($\gamma _{R}=c_{2N}$) which are completely localized on the leftmost (rightmost) Majorana site $1$ ($2N$), describing \textquotedblleft half\textquotedblright\ a fermion each. These edge modes exist in the whole parameter regime $-2J<\mu <2J$, however leaking more and more strongly into the wire when approaching the critical values. Their existence is robust against perturbations such as disorder, which can be traced back to the bulk gap in connection with their topological origin \cite{Kitaev00}. 

\begin{figure}[t]
\centering
\includegraphics[width=7.5cm]{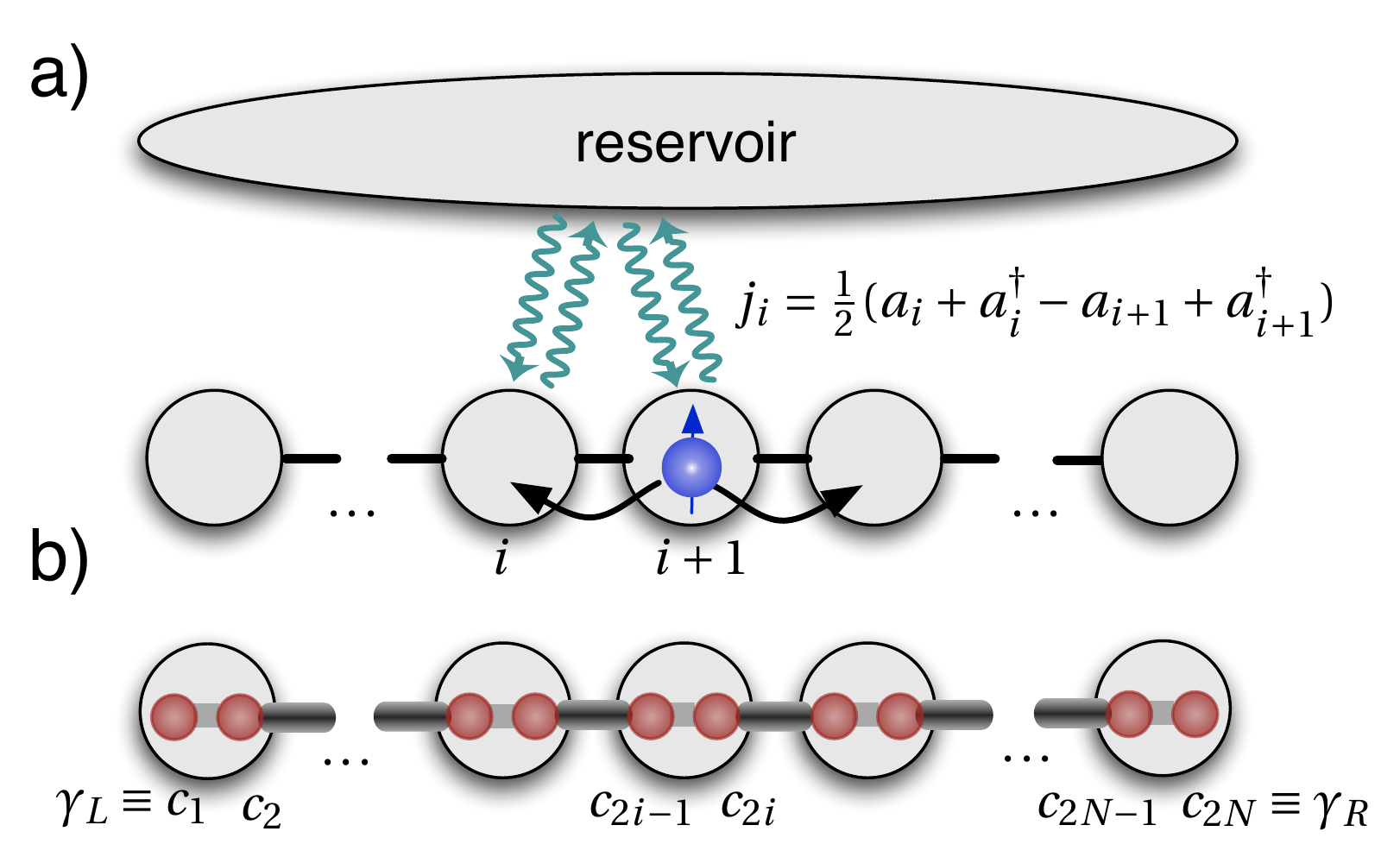} 
\begin{caption}{\label{Fig1_Setup}Schematic setup for the dissipative Majorana quantum wire. a)  A reservoir represents a source and drain for the quantum wire which is coherent over each pair of sites. Independent of the initial condition, the bulk of the system is then cooled into a p-wave superfluid state by dissipatively establishing a pairing link for each two adjacent sites. b) Illustration of the stationary state, where in the Majorana basis of real fermions each physical site is split into two Majorana sites. In the bulk all Majorana modes from neighboring sites are paired (black links). For a finite wire two dissipative unpaired Majorana modes $\gamma_L,\gamma_R$ appear at the edge as a highly nonlocal decoherence free subspace. They are isolated from the bulk by a dissipative gap.  For a realization with cold atoms, see Fig.~\ref{fig2}. }
\end{caption}
\end{figure}

\emph{Dissipative topological quantum wire} -- Consider the master equation for spinless fermions in a 1D chain with $N$ sites, 
\begin{equation}\label{equ:ME}
\partial _{t}\rho =-\mathrm{i}\,[H,\rho ]+\kappa \sum_{i}\left[ j_{i}\rho j_{i}^{\dag }-\tfrac{1}{2}\{j_{i}^{\dag }j_{i},\rho \}\right] \equiv \mathcal{L}[\rho ],
\end{equation}
with $\rho $ the system density operator. The two terms on the right hand side are a Hamiltonian and a dissipative term, respectively. Here we concentrate on purely dissipative dynamics which occurs at rate $\kappa $, and set $H=0$. We choose the Lindblad operators $j_{i}$ as the Bogoliubov operators defined above, 
\begin{equation}\label{equ:jump}
j_{i}\equiv \tilde{a}_{i}=\tfrac{1}{2}(a_{i}+a_{i}^{\dag}-a_{i+1}+a_{i+1}^{\dag }), ~\, ~ (i=1,\ldots ,N-1).
\end{equation}
These Lindblad operators are quasi-local superpositions of annihilation and creation operators (see Fig.~\ref{Fig1_Setup}). Due to the fermionic nature of $j_i$, this choice ensures that the bulk of the system \textquotedblleft cools\textquotedblright\ under the above dynamics to the unique\emph{\ pure state} $\tilde{a}_{i}|G\rangle =0$, which by construction agrees with the $p$-wave superfluid ground state of the Hamiltonian (\ref{HKit}). Following \cite{Diehl08,Verstraete09,Diehl10b}, this steady state is thus a \emph{many-body dark state} of the Liouvillian, $\mathcal{L} \left( |G\rangle \langle G|\right) =0$.

The approach to this steady state is governed by the damping spectrum of the Liouvillian $\mathcal{L}$. Diagonality of $\mathcal{L}$ in the $\tilde{a}_{i} $ implies a flat damping spectrum $\kappa _{k}=\kappa $ in analogy to the excitation spectrum of the Hamiltonian above. While the damping spectrum $\kappa _{k}\geq 0$ is always positive semi-definite for fermions, this \textquotedblleft dissipative gap\textquotedblright\ $\kappa _{0}\equiv \min (\kappa _{k})=\kappa $ implies exponentially fast approach of all observables to their steady state values.

For a finite wire we find dissipative zero modes related to the absence of the Lindblad operator $\tilde{a}_{N}$. More precisely, there exists a subspace spanned by the edge-localized Majorana modes $\tilde{a}_{N}=\frac{1}{2} \left(\mathrm i\gamma _{L}+\gamma _{R}\right) $, with the above Fock basis $|\alpha \rangle \in \{ |0\rangle ,|1\rangle \}$, which is decoupled from dissipation, i.e. $\partial _{t}\rho_{\alpha \beta }(t)=0$ with $\rho _{\alpha \beta }\equiv \langle \alpha |\rho |\beta \rangle $.  

\emph{Edge modes as nonlocal decoherence free subspace} -- These dissipative edge modes are readily revealed in solutions of the master equation.  Eq.~(\ref{equ:ME})  is quadratic in the fermion operators, which implies solutions in terms of Gaussian density operators $\rho(t) \sim \exp \left[ -\tfrac{\mathrm i}{4} c^{T}G(t)c \right] $. Here we have defined a column vector $c$ of the $2N$ Majorana operators, and $G$ is a real antisymmetric matrix related to the correlation matrix $\Gamma _{ab}(t)=\tfrac{\mathrm{i}}{2}\langle [c_{a}, c_{b}]\rangle = \mathrm i [\tanh (\mathrm i G/2)]_{ab}$, which equally is real antisymmetric. Writing the Lindblad operators in the Majorana basis, $j_i =l_{i}^Tc, j^\dag _i = c^T l_{i}^*$, such that the Liouvillian parameters are encoded in a hermitian $2N\times 2N$ matrix $M = \sum_i  l_{i}\otimes l_i^\dag $, this covariance matrix is seen to obey the dissipation-fluctuation equation \cite{Eisert11} 
\begin{equation}\label{equ:FD}
\partial _{t}\Gamma =- \{X,\Gamma \} -Y,
\end{equation}
with real matrices $X=2 \mathrm{Re} M = X^T$ and $Y = 4 \mathrm{Im} M  = -Y^T$. Physically, the matrix $X$ describes a drift or damping, while the matrix $Y$ is related to fluctuations in steady state, determined by $\{X,\bar{\Gamma}\} =  - Y$. Writing $\Gamma = \bar{\Gamma}+\delta\Gamma$, the approach to steady state is governed by $\partial_t \delta \Gamma = - \{X,\delta\Gamma\}$, i.e., the eigenvalues of the positive semi-definite matrix $X$ \cite{Prosen08} give the damping spectrum. The ``dark'' nonlocal subspace of edge modes, decoupled from dissipation, is thus associated with the subspace of zero eigenvalues of the damping matrix $X$. In a spectral decomposition $X = \sum_r \lambda_r |r \rangle \langle r|$, and identifying by greek subscripts the zero eigenvalues subspace, we can write by partitioning  
\begin{eqnarray}\label{Partition}
\partial_t\left(\begin{array}{cc}
\Gamma_{\alpha\beta}  & \Gamma_{\alpha s}\\\vspace{0.1cm}
\Gamma_{r\beta} & \Gamma_{rs} 
\end{array} \right)
&=& \left( 
\begin{array}{cc}
0 & - ( \Gamma \lambda )_{\alpha s}   \\\vspace{0.1cm}
- (\lambda \Gamma)_{r\beta}  & ( - \{\lambda,\Gamma\} -  Y) _{rs} 
\end{array} 
\right).
\end{eqnarray}
While the bulk ($rs$ sector) damps out to the steady state by dissipative evolution, the density matrix in the edge mode subspace ($\alpha\beta$ sector) does not evolve, preserving its initial correlations. The coupling density matrix elements (mixed sectors) damp out according to $\Gamma_{r\beta}(t)= e^{- \lambda_r t} \Gamma_{r\beta}(0)$; in the presence of a dissipative gap as in the example above, this fadeout of correlations is exponentially fast, leading to a dynamical decoupling of the edge subspace and the bulk. More generally, this structure of the master equation appears whenever there exists a basis in which each Lindblad operator is block diagonal with blocks associated to edge and bulk, and with vanishing entries in the edge block (see appendix).

\begin{figure}[t] \centering
\includegraphics[width=8.5cm]{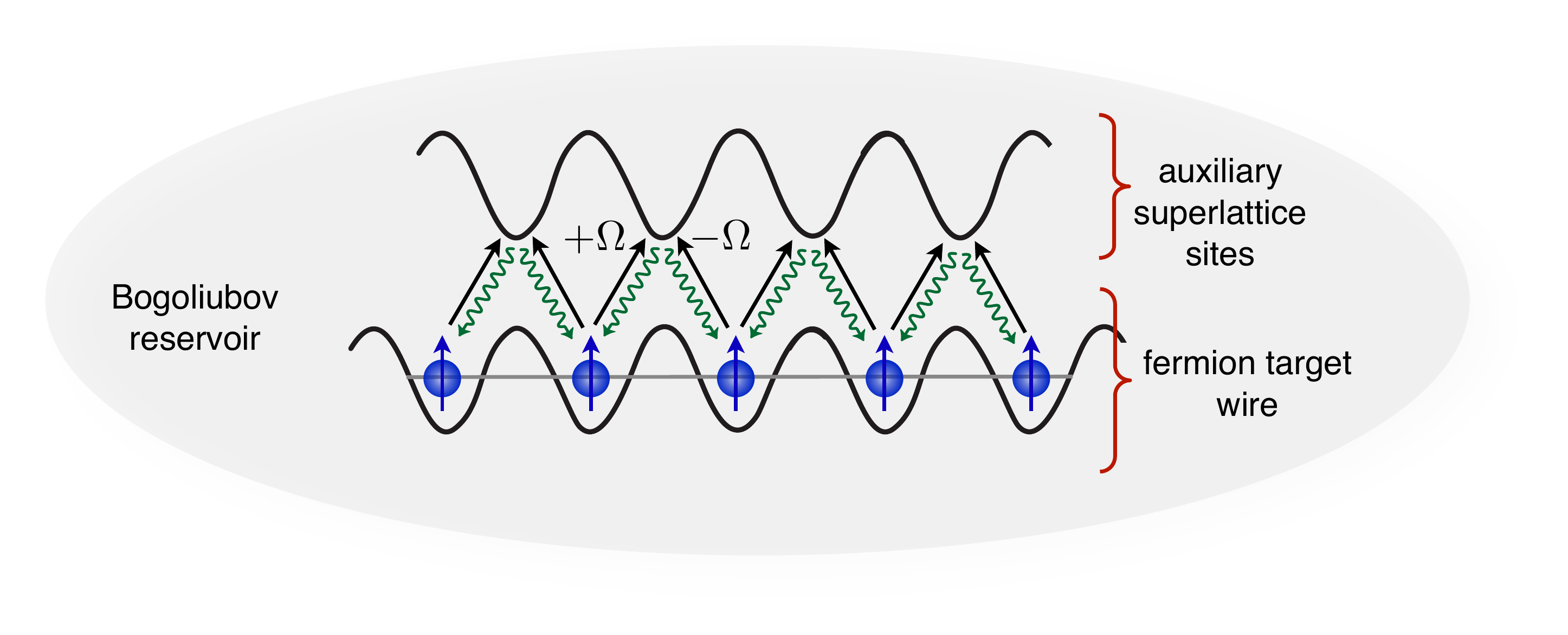}
\begin{caption}{\label{ImplementationMicro}Microscopic implementation scheme for the Majorana Liouvillian Eqs. (\ref{equ:ME},\ref{equ:jump}). The quantum wire is represented by the lower sites of an optical superlattice for spin polarized atomic fermions. They are coherently coupled to auxiliary upper sites by lasers with Rabi frequencies $\pm\Omega$, alternating from site to site. Dissipation results from spontaneous Bogoliubov phonon emission via coupling of the system to a BEC reservoir (light grey). An edge can be created using single site addressability tools \cite{Greiner10,Kuhr10}, cutting off the lattice at an auxiliary site instead of a target system site. As shown in the implementation section, this setting reduces to the Lindblad operators (\ref{equ:jump}) at late times. \label{fig2} }
\end{caption}
\end{figure}

In summary, we arrive at the physical picture that dissipative evolution cools the bulk into a p-wave superfluid and thereby isolates the edge mode subspace, $\rho (t\rightarrow \infty )\rightarrow \rho _{\text{edge}}\otimes \rho _{\text{bulk}}$, providing a highly nonlocal decoherence free subspace \cite{Lidar98}. A physical implementation of the  master equation (\ref{equ:ME}) with cold atoms is outlined schematically in Fig.~\ref{fig2}. More details on the setup are given in the implementation section, following ideas of Ref.~\cite{Diehl08}. 

\section{Stability of edge mode subspace}

Here we study the robustness of the edge mode subspace against (i) global parameter changes in the Lindblad operators, while preserving their translation invariance, and (ii) static disorder, which breaks this invariance. We consider two examples of quadratic master equations (\ref{equ:ME}) with Lindblad operators deviating from the ideal case (\ref{equ:jump}),
\begin{eqnarray}\label{deformations}
j^{(c)}_i = \tfrac{1}{\sqrt{2}} (  \sin{\theta} \, (  a_i - a_{i+1} )+ \cos{\theta} \, ( a^{\dagger}_i  + a^{\dagger}_{i+1} ) ),\\
j^{(n)}_i = \tfrac{1}{\sqrt{2}} (  \sin{\theta} \, ( a^{\dagger}_i -  a_{i+1} ) + \cos{\theta} \, (  a_i +a^{\dagger}_{i+1}  ) ),\label{deformations1}
\end{eqnarray} 
where the ideal case corresponds to $\theta = \pi/4$. In the first case, the steady state of the bulk remains pure. In the second case, we find a mixed state while still preserving the properties of the edge subspace. As elaborated on in the appendix, this results from the fact that the first case (c) represents a canonical transformation up to normalization of the ideal Lindblad operators in momentum space,
 while the second one (n) is not (cf. Eq. (\ref{quasicanonical}) in the appendix).
 In this latter case, the steady state has no counterpart as a ground state of some Hamiltonian. This difference is illustrated in Figs.~\ref{fig2} a), b). There, we plot the purity spectrum $\mathrm{spec}\,( \bar \Gamma^2)$ in steady state (with the edge mode subspace initialized as pure); a pure state in the bulk is indicated by all eigenvalues of $\bar \Gamma^2$ being equal to $-1$. Note that static disorder, implemented in terms of small random variations of the Lindblad parameters of range $\epsilon\ll\theta$ from site to site, makes the first case non-equivalent to a canonical transformation, and, therefore, degrades the purity of the steady state.

\begin{figure}[t] \centering
\includegraphics[width=8.5cm]{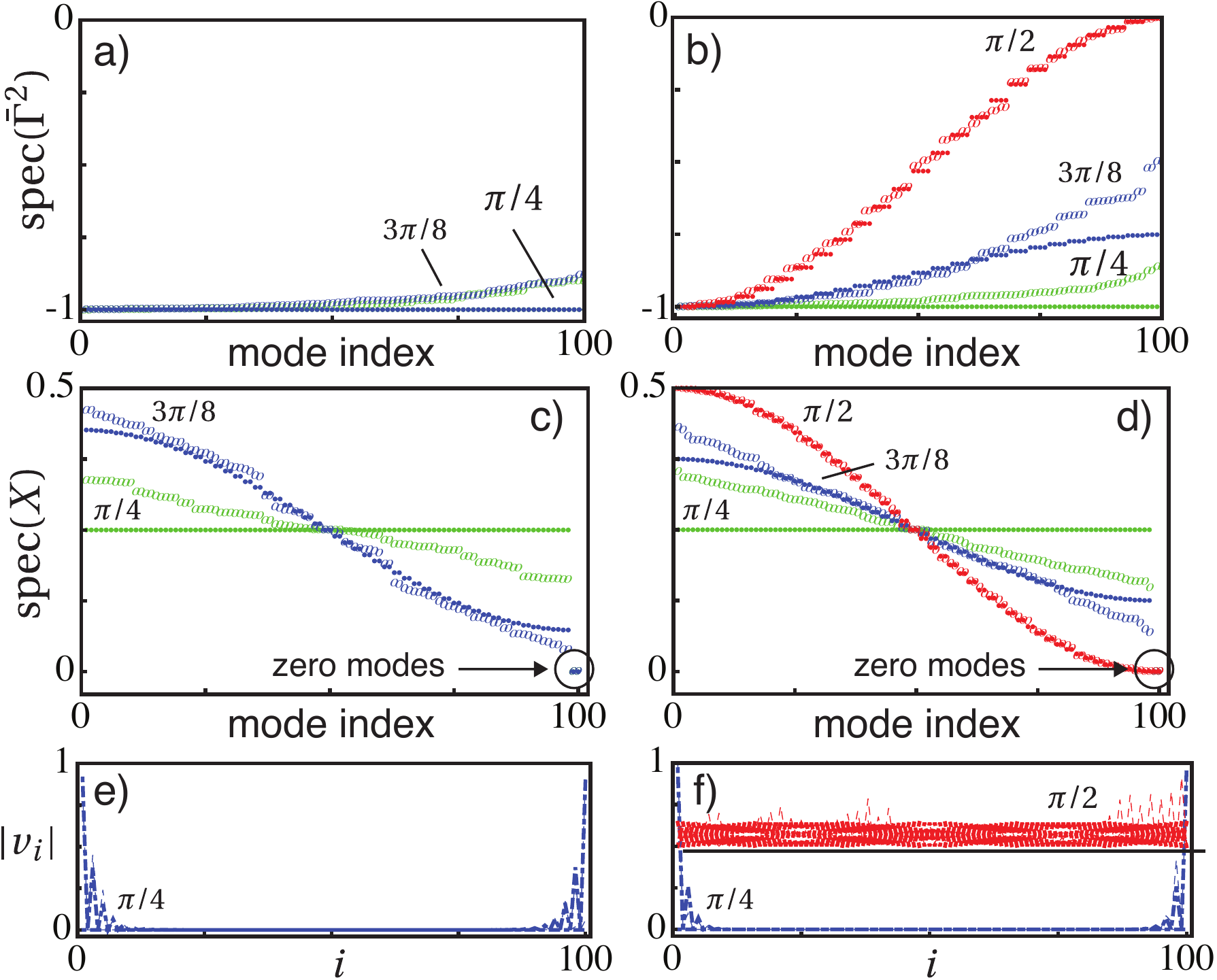}
\begin{caption}{\label{fig3} For a quantum wire with $50$ lattice sites we plot  the eigenvalue spectrum of the square steady state correlation matrix ${\bar \Gamma}^{2}$ (Figs. a) and b)), and the eigenvalue spectrum of the damping matrix $X$ (Figs. c) and d)) as a function of the eigenvalue index. In the last row, the amplitude moduli  $|v_{i}|$ for the left and right zero mode (edge mode) as a function of the lattice site index $i$ are shown (Figs. e) and f)). The first and second column corresponds to quasi-canonical (\protect\ref{deformations}) and non-canonical Lindblad operators (\protect\ref{deformations1}), respectively. Results are shown for three angles: The ideal case $\theta=\frac{\pi}{4} $ Eq.~(\ref{equ:jump})  (green), $\theta=\frac{3\pi}{8}$ (blue) and $\theta=\frac{\pi}{2}$ (red). Dissipative zero modes exist in all cases, but they become degenerate with the bulk modes in the third case (red) case where the bulk dissipative gap collapses. The closed circles show results for a globally fixed $\theta$, while the open circles correspond to addition of local static disorder to the angles. 
}
\end{caption}
\end{figure}

While the purity spectrum is  qualitatively different for both kinds of parameter deformations, the spectra of damping matrices, $\mathrm{spec}\,(X)$, are rather similar. The characteristic features are (i) the existence of a dissipative gap $\kappa_0 = \kappa\cos\theta$, closing at $\theta = \pi/2$, and (ii) the existence of two zero modes throughout the parameter space. The associated orthogonal eigenvectors $v_{L,R}$ describing the Majorana modes $\gamma_{L,R} = v_{L,R}^T c$ can be constructed explicitly (see appendix), with localization length given by $l_\text{loc}/a= (\log  | \tfrac{\sin \theta + \cos\theta}{\sin \theta - \cos\theta}|)^{-1}$ in both cases, in units of the lattice constant $a$. This shows the characteristic exponential edge localization of Majorana modes close to the ideal case, while their extent becomes comparable to the system size close to the gap closing points (see Fig. \ref{fig3}). Adding disorder modifies the bulk spectrum quantitatively of $\mathcal O(\epsilon)$, while the zero mode subspace persists. In fact, the existence of two zero eigenvalues in the presence of disorder can be established by explicit construction (see appendix). 

In summary, a two-dimensional decoherence free subspace exists for the entire parameter range notwithstanding disorder or the fact that the bulk steady state may be  strongly mixed. A sensible notion of protection of the subspace in terms of dissipative isolation from the bulk, however, only exists sufficiently far away from the point where the damping gap closes.

\section{Adiabatic parameter changes and dissipative braiding}

The above can be generalized to a dissipative quantum wire network of $M$ finite chains, as counterpart of the Hamiltonian networks discussed by Alicea {\em et al.}~\cite{Alicea11}. This results in higher dimensional non-evolving subspaces of dimension $2M$, again governed by the structure given by Eq.~(\ref{Partition}). As we will show below, this leads to the possibility of braiding of the dissipative Majorana edge modes by adiabatic parameter changes in the Liouvillian. 

We consider the time evolution of the density matrix in a co-moving basis $|a(t) \ra = U(t) | a(0) \ra$ which follows the decoherence free subspace of edge modes, i.e.~preserves the property $\dot{\rho}_{\alpha \beta}= 0$. Demanding normalization of the instantaneous basis for all times, $\la b(t) |a(t) \ra = \delta_{ab}$, this yields
\be
\frac{d}{dt} \rho = -\mathrm i [A , \rho ] + \sum_{a,b} |a\ra \dot{\rho}_{ab} \la b |,
\ee
with the unitary connection operator $A =\mathrm i \dot{U}^{\dagger} U$ and $\dot{\rho}_{ab} \equiv \la a(t) | \partial_t \rho | b(t) \ra $ the time evolution in the instantaneous basis. The Heisenberg commutator clearly reflects the emergence of a gauge structure \cite{Berry84,Simon83,WilczekZee83,Carollo2003a,Pachos99} in the density matrix formalism, which appears \emph{independently} of what kind of dynamics -- unitary or dissipative -- generates the physical time evolution, represented by the second contribution to the above equation. The transformation exerted on the zero mode subspace of either Hamiltonian or Liouvillian with an initial condition $\rho_{\alpha \beta}(0)$ is then given by $\rho_{\alpha \beta}(t) = \left( V(t) \rho(0) V(t)^{\dagger} \right)_{\alpha \beta}$, with time-ordered $V(t) = T\exp{\left( -\mathrm i \int^t_0 d\tau A(\tau)  \right)}$ and $A(t)_{\alpha \beta} =\mathrm i \la \dot{ \alpha}(t) | \beta(t) \ra$.

Of central importance for such state transformations to work without losing the protected subspaces is adiabaticity of the parameter changes. Here, this is a requirement on the ratio of the rate of parameter changes $\dot{\theta}$ versus the bulk dissipative gap $\kappa_0$. This separation of time scales, naturally provided due to the non-evolving subspace, prevents the protected decoherence-free subspace from ever being left, a phenomenon sometimes referred to as the Quantum Zeno effect \cite{Beige00}. 

\begin{figure}[t]
\centering
\includegraphics[width=8.5cm]{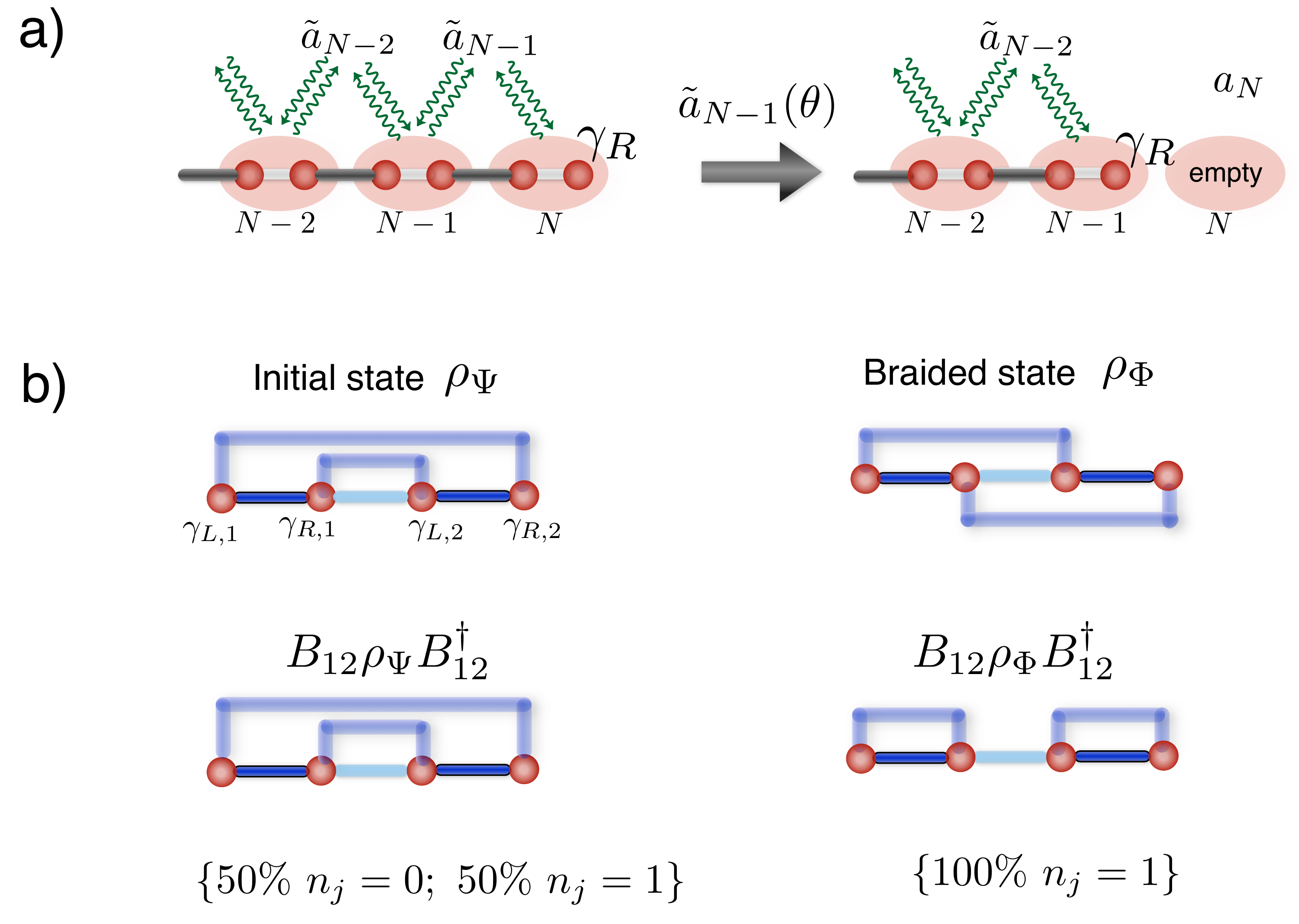}
\begin{caption}{\label{Fig3_MajMove} a) Elementary Majorana move: transfer of a Majorana edge mode one site to the left by adiabatic change of the parameter $\theta$ in the dissipative Liouville operator. b) Illustration of the initial state $\rho_\Psi$, braided state $\rho_{\Phi}$ (upper row), the same states in the changed basis together with the results of the fermionic number measurements in the interferometric gedankenexperiment. The non-zero correlations between Majorana fermions are shown as semitransparent lines.}
\end{caption}
\end{figure}

Equipped with this understanding of the role of the dissipative gap, we now
demonstrate how to move a Majorana fermion adiabatically along the wire in our
dissipative setup (cf.~Fig.~\ref{Fig3_MajMove} a)), which is the key ingredient
to perform braiding. As an example, we describe the move of the right unpaired
Majorana fermion $\gamma_{R}$ from site $N$ ($\gamma_{R}=c_{2N}$) to site
$N-1$ ($\gamma_{R}=c_{2N-2}$). To achieve this purpose, we consider an
adiabatic change of the last Lindblad operator $\tilde{a}_{N-1} $ of the
form: $\tilde{a}_{N-1}(\theta)=\frac{1}{2} [  a_{N}^{\dagger}-a_{N}%
+\cos{\theta}(a_{N-1}^{\dagger}+a_{N-1})-\sin{\theta}(a_{N}^{\dagger}%
+a_{N}) ]  $, where $\theta$ adiabatically varies with time from
$\theta=0$ where $\tilde{a}_{N-1}(0)=\tilde{a}_{N-1}$ to $\theta=\pi/2$ where
$\tilde{a}_{N-1}(\pi/2)=-a_{N}$. At the end of this evolution, the site $N$ is
empty (vacuum), and the right Majorana fermion moves one site to the left.
This is the analog of locally tuning the chemical potential in the Hamiltonian
setting to move Majoranas \cite{Alicea11}. For the evolution of the Majorana
mode population induced by a finite ramping velocity for time $T$, we find
$\la c_{2N-2} c_1 \ra=\la c_{2N} c_1 \ra \exp{\left(  -2\int_{0}^{T}dt\dot{\theta} %
^{2} / \kappa \right)  }$, describing a weak dephasing of the Majorana mode (see appendix).

Operating this mechanism on a T-junction \cite{Alicea11} in order to exchange the two modes adiabatically while permanently keeping them sufficiently far apart from each other, the unitary braiding matrix describing the process is $B_{ij}=\exp{\left(  \frac{\pi}{4} \gamma_{i}\gamma_{j} \right)  }$ for two Majorana modes $i,j$, demonstrating non-abelian statistics since
$[B_{ij},B_{jk}]\neq0$ for $i\neq j$. While knowledge of the braiding
matrix for each pair of Majorana modes is in principle enough to demonstrate
non-abelian statistics, braiding of a single pair of Majorana modes is not
physically observable since super-selection rules dictate a diagonal density
matrix for a single complex fermion, which therefore is insensitive to the
acquired phase.

We follow \cite{Bravyi,Freedman} to construct an
interferometric experiment that explicitly shows the non-abelian nature of
dissipative braiding of Majoranas fermions. The setup, cf.~Fig.~\ref{Fig3_MajMove} b), is given by two finite
wires which host four unpaired Majorana modes, $\gamma_{L,1}$ and
$\gamma_{R,1}$ for the left and right unpaired Majorana fermions of the first
wire, and similarly $\gamma_{L,2}$ and $\gamma_{R,2}$ for the second one. These
four real fermions correspond to two complex fermions $a_{j}=(\gamma
_{R,j}+\mathrm{i} \gamma_{L,j})/2$ with $j=1,2$. For the subspace with an even number of
complex fermions, we define the following basis $|\bar{0}%
\ra=|00\ra=|\text{vac}\ra,$ and $|\bar{1}\ra=|11\ra=a_{1}^{\dagger}%
a_{2}^{\dagger}|\text{vac}\ra$. We now prepare an intial state of the system
as $|\Psi\ra=\left(  |\bar{0}\ra-|\bar{1}\ra\right)  /\sqrt{2}$ (the
corresponding density matrix is $\rho_{\Psi}=|\Psi\ra\la\Psi|$) such that
$\la\gamma_{R,1}\gamma_{L,2}\ra=\la\gamma_{R,2}\gamma_{L,1}\ra=\mathrm{i}$. If we now
braid the Majoranas $\gamma_{R,1}$ and $\gamma_{L,1}$, the initial state
$\rho_{\Psi}$ transforms into $\rho_{\Phi}=B\rho_{\Psi}B^{\dagger}$, where
$B=\exp{( \frac{\pi}{4}\gamma_{L,1}\gamma_{R,1})} $. To distinguish the states $\rho_{\Psi}$ and $\rho_{\Phi}$ in an occupation number measurement, we first make a unitary change of basis via
$B_{12}=\exp{(\frac{\pi}{4}\gamma_{R,1}\gamma_{L,2} )}$ such that %
\be
\begin{split}
B_{12}\rho_{\Psi}B_{12}^{\dagger} &  =\tfrac{1}{2}\left(  |\bar{0}\ra-|\bar
{1}\ra\right)  \left(  \la\bar{0}|-\la\bar{1}|\right)  =\rho_{\Psi},\\
B_{12}\rho_{\Phi}B_{12}^{\dagger} &  =|\bar{1}\ra\la\bar{1}|.
\end{split}
\ee
and then measure the number of fermions $n_{1}=\la a_{1}^{\dagger}a_{1}\ra$
and $n_{2}=\la a_{2}^{\dagger}a_{2}\ra$ on the wires. In the first case, we
get with equal probabilities either $n_{j}=0$ or $n_{j}=1$, while we always
get $n_{j}=1$ in the second case, cf. Fig. \ref{Fig3_MajMove} b).


\section{Topological Order of the Steady State}

We now show that the robustness of the edge modes seen above is indeed related to the
existence of topological order in the bulk of a stationary state of Liouvillian evolution,
and construct a topological invariant, which characterizes topologically different states.
This classification does not
rely on the existence of a Hamiltonian or on the purity of a state
(in contrast to existing constructions involving ground states of Hamiltonians), and can be
entirely formulated in terms of a density matrix. 

In an infinite system, a stationary state of a
Gaussian translationally invariant Liouvillian is described by a
density matrix $\rho=\prod_{k\geq0}\rho_{k}$, where $\rho_{k}$ is a $4\times 4$
hermitian unit trace matrix, which describes the momentum mode pair $\pm k$. The
topologically relevant information is encoded in the $2\times 2$ block
$\rho_{2k}$ of the density matrix in the subspace with even occupation of the
modes $\pm k$, $\langle a_k^\dag a_k\rangle+ \langle a_{-k}^\dag a_{-k}\rangle=0,2$ (see appendix). The matrix $\rho_{2k}$ is proportional to
$\tfrac{1}{2}(\mathbf{1}+\vec{n}_{k}\vec{\sigma})$, where $\vec{\sigma}$ is
the vector of Pauli matrices and $\vec{n}_{k}$ is a real three-component
vector $0\leq|\vec{n}_{k}|\leq1$. The pure states correspond to $\rho_{k}%
^{2}=\rho_{k}$, i.e. $|\vec{n}_{k}|=1$ for all $k\geq0$. We can naturally
extend the definition of $\vec{n}_{k}$ to negative $k$ following the change of
$\rho_{2k}$ resulting from the transformation of the basis vectors in the subspace
under $k\rightarrow-k$: $n_{-k}^{x,y}\rightarrow-n_{k}^{x,y}$ and $n_{-k}%
^{z}\rightarrow n_{k}^{z}$. Note that the vector $\vec{n}_{k}$ is
continuous at $k = 0,\pm \pi$ because $\vec{n}_{k=0}$ and $\vec{n}%
_{k=\pm\pi}$ have only $z$-component.

\begin{figure}[t] \centering
\includegraphics[width=9.0cm]{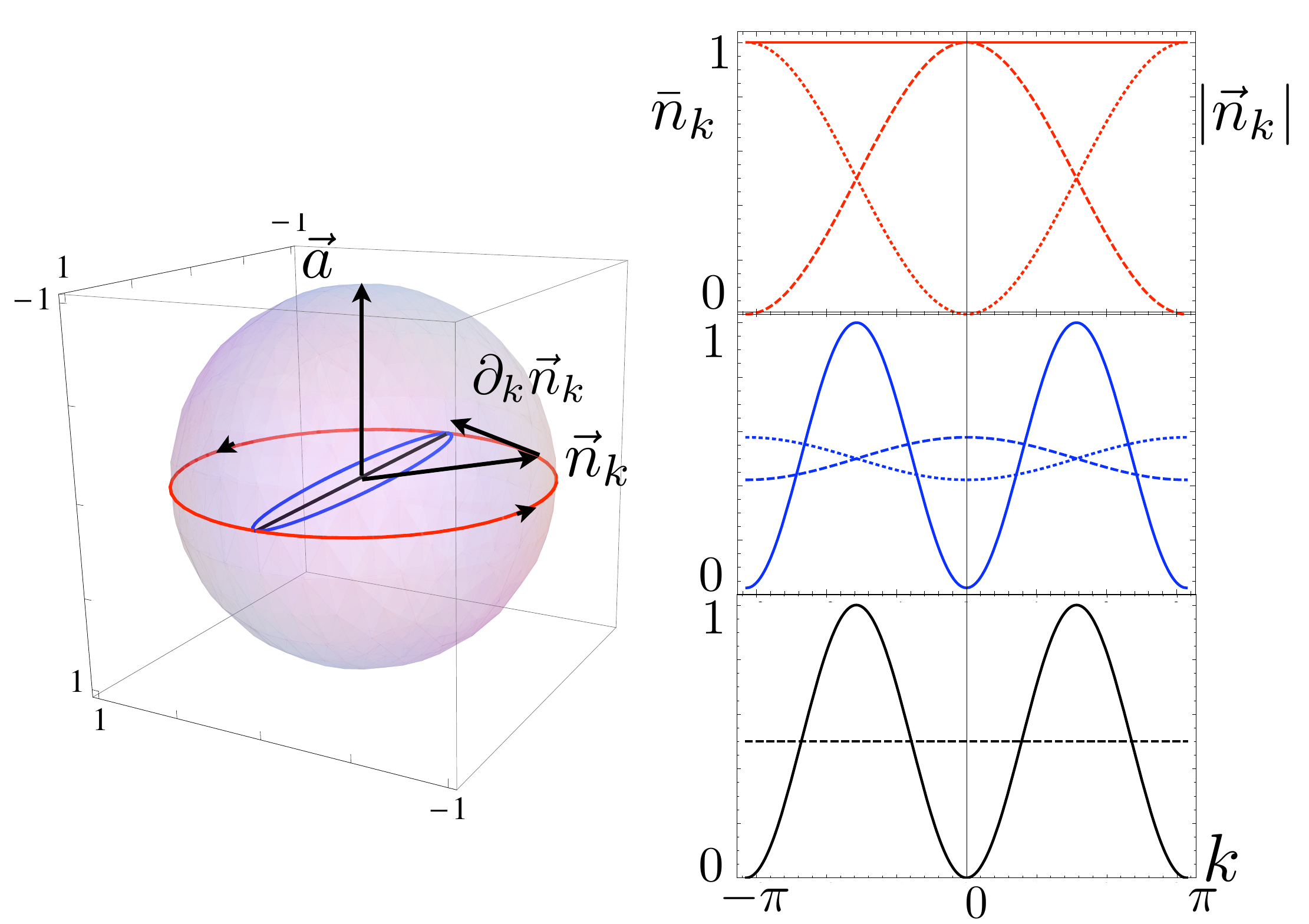}
\begin{caption}{\label{Fig4_TopInv} Visualization of the topological invariant $\nu$ for chirally symmetric mixed states. Left panel: Chiral symmetry constrains $\vec n_k$ to a great circle. For pure states, it is furthermore pinned to unit length (red). Tuning the Liouville parameters destroys the purity and deforms the circle to an ellipse (blue). A phase transition occurs when the ellipse shrinks to a line (black). Crossing the transition, the topological invariant changes sign from $+1$ to $-1$. Right panel:  Purity $|\vec n_k(\theta)|$ (solid) and average occupation $\bar n_k(\theta) = \tfrac{1}{2} (1 - n_{z,k}(\theta))$ (dashed and dotted) for various fixed transition parameter $\theta$ for both sides of the transition (see text). Throughout the transition, the system is half filled. }
\end{caption}
\end{figure}

Once the vector $\vec{n}_{k}$ is nonzero for all $k$, the normalized vector
$\hat{\vec{n}}_{k}=|\vec{n}_{k}|^{-1}\vec{n}_{k}$ defines a mapping
$S^{1}\rightarrow S^{2}$ of a circle $S^{1}$ (the Brillouin zone $-\pi\leq
k\leq\pi$ with identified end points $k=\pm\pi$ due to usual periodicity in the
reciprocal lattice) into a unit sphere $S^{2}$ of end points of $\hat{\vec{n}%
}_{k}$. This mapping, however, is topologically trivial (the corresponding
homotopy group $\pi_{1}(S^{2})=0$), since a circle can always be continuously
deformed into a point on the sphere. We therefore need an additional
constraint on $\vec{n}_{k}$ in order to introduce a nontrivial topology. In
our setting, motivated by Kitaev's model Hamiltonian \cite{Kitaev00}, the
constraint is provided by the chiral symmetry \cite{Altland97,Ryu10}. In terms
of the density matrix, the chiral symmetry is equivalent to the existence of a
$k$-independent unitary matrix $\Sigma$ with $\Sigma^{2}=\mathbf{1}$, which
\textit{anticommutes} with the traceless part of the density matrix ($\vec
{n}_{k}\vec{\sigma}$ in our case): $\Sigma\,\vec{n}_{k}\vec{\sigma}%
\,\Sigma=-\vec{n}_{k}\vec{\sigma}$. After representing the matrix $\Sigma$ in
the form $\Sigma=\vec{a}\vec{\sigma}$, where $\vec{a}$ is a constant unit
vector, the chiral symmetry condition reads $\vec{n}_{k}\vec{a}=0$, i.e., the
vector $\vec{n}_{k}$ is orthogonal to $\vec{a}$ for all $k$. The end point of
$\hat{\vec{n}}_{k}$ is now pinned to a great circle $S^{1}$ on the sphere such
that the vector $\hat{\vec{n}}_{k}$ defines a mapping $S^{1}\rightarrow S^{1}$
from the Brillouin zone into a circle. The corresponding homotopy group is now
nontrivial, $\pi_{1}(S^{1})=\mathbf{Z}$, and such mappings are divided into
different topological classes distinguished by an integer topological invariant
(winding number). The explicit form of this invariant reads (see appendix)%
\begin{eqnarray}\label{winding}
\nu=\frac{1}{2\pi}\int_{-\pi}^{\pi}dk\,\vec{a}\cdot(\hat{\vec{n}}_{k}%
\times\partial_{k}\hat{\vec{n}}_{k})\in\mathbf{Z}.
\end{eqnarray}
Geometrically, $\nu$ counts the number of times the unit vector $\hat{\vec{n}%
}_{k}$ winds around the origin when $k$ goes across the Brillouin zone, cf.
Fig. \ref{Fig4_TopInv}. Importantly, the winding number $\nu$
distinguishes topologically different density matrices for translationally
invariant Gaussian systems with chiral symmetry without restriction on the purity
of the state.

Let us now discuss specific examples of the steady states resulting from
quasi-local dissipative dynamics. As shown in the appendix, for momentum space
Lindblad operators $j_{k}=\xi_{k}^{T}\Psi_{k}$, $\xi_{k}^{T}=(u_{k}%
,v_{k}),\Psi_{k}^{T}=(a_{k},a_{-k}^{\dag})$ with (unnormalized) Bogoliubov
functions $u_{k},v_{k}$ which do not induce a current in the steady state
($\langle a_k^\dag a_k\rangle= \langle a_{-k}^\dag a_{-k}\rangle$), the solution for the vector $\vec{n}_{k}$ reads
\begin{eqnarray}
\vec{n}_{k}=\frac{1}{2}\left(
\begin{array}
[c]{c}%
m_{x,k}-m_{x,-k}\\
m_{y,k}-m_{y,-k}\\
m_{z,k}+m_{z,-k}%
\end{array}
\right)  ,\quad\vec{m}_{k}=\kappa_{k}^{-1}\xi_{k}^{\dag}\vec{\sigma}\xi_{k},
\end{eqnarray}
with $\kappa_{k}=(\xi_{k}^{\dag}\xi_{k}+\xi_{-k}^{\dag}\xi_{-k})/2$. For
quasi-canonical deformations (cf. Eq. (\ref{deformations})), the additional symmetry
property $u_{-k}=u_{k}$ and $v_{-k}=-v_{k}$ simplifies the solution to
$\vec{n}_{k}=\vec{m}_{k}$, implying the purity of the steady state, $|\vec
{n}_{k}|=1$ for all $k$. The solution corresponds to the ground state of some
Hamiltonian. For $\theta=\theta_{s}=\pi s/2$ with an integer $s$, the system
has a damping gap closing point, and the vector $\vec{n}_{k}$ has only
$z$-component ($-1$ for even $s$ and $+1$ for odd $s$) for all $k$, leading to
$\nu=0$. For other values of $\theta$, one has $\nu=\pm1$. However, as one
cannot define the direction of $\vec{a}$ for $\theta_{s}$, a global definition
of $\nu$ for all $\theta$ is not possible, with the consequence that the potential change of the
sign of $\nu$ when $\theta$ passes $\theta_{s}$ is meaningless. Note that
the gap closing points do not correspond to a phase transition here, because one
has either a completely empty or completely filled lattice, such that no
thermodynamic observables of a phase transition could sensibly be defined.

For non-canonical deformations, Eq. (\ref{deformations1}), visualized in Fig.
\ref{Fig4_TopInv}, the steady state density matrix is mixed, $|\vec{n}%
_{k}|\leq1$ (cf. also right panel in Fig. \ref{Fig4_TopInv}). Importantly, the
topological order persists for quite strongly mixed states, and we find again
$\nu=\pm1$ for $\theta\neq\theta_{s}$. The difference to the previous
example is that at $\theta=\theta_{s}$, not only the direction of $\vec{a}$ but
also the topological invariant is not defined: $\vec{n}_{k}$, alined in the
$y$-direction for all $k$, has zeroes: $\vec{n}_{k=0,\pi}=0$, meaning physically that these modes are in a completely mixed state. The ``loss'' of
topology at $\theta=\theta_{s}$ can be viewed as a non-equilibrium
topological phase transition \cite{Rudner09,Lindner11,Kitagawa10} as a
result of changing the Liouville parameters: The system has well defined
thermodynamic properties, since it is half filled, $\bar{n}(\theta)=\int
\tfrac{dk}{2\pi}\bar{n}_{k}(\theta)=1/2$ for all $\theta$. The closing of the
dissipative gap at $\theta_{s}$ leads to critical behavior, which manifests
itself via diverging time scales, resulting e.g  in an algebraic approach to steady
state (as opposed to exponential behavior away from criticality)
\cite{Diehl08,Verstraete09,Diehl10a,Eisert11}. Importantly, the vector $\vec
{n}_{k}$ in the steady state has the reflection property $\vec{n}_{k}(-\delta
\theta)=P\vec{n}_{k}(+\delta\theta)$ for all $k$, where $P=\mathrm{diag}(1,1,-1)$ and
$\delta\theta=\theta-\theta_{s}$. Therefore, the symmetry
pattern of the steady state is identical on both sides of the transition,
ruling out a conventional Landau-Ginzburg type transition and underpinning the
topological nature of the transition.

We see that in our system the stationary state of Liouvillian evolution in an infinite system is indeed characterized 
(for $\theta \neq \theta_{s}$) by a nontrivial topological order. If the system is finite, the edges
separate the topologically non-trivial bulk from a vacuum, which is topologically trivial.
Similar to the Hamiltonian case of topological insulators or superconductors \cite{HasanKane},
the ``jump'' in the topological order guarantees the existence of edge states. For a pure
stationary state (quasi-canonical case)
this follows from the formal analogy with the Hamiltonian case in which it is well-established (see e.g.~Ref.~\cite{gurarie11}). 
More generally, for a mixed state (non-canonical case), the arguments follow the line of Ref.~\cite{Kitaev00} using an alternative equivalent form of Eq. (\ref{winding}) for the topological invariant discussed in the appendix, cf.~Eq.~(\ref{windingVec}). The topological origin of the edge modes explains their
stability demonstrated above.

\section{Implementation in Cold Atomic Gases}  

Here we devise an interacting (quartic in the fermion operators) Liouville operator which at late times reduces to the quadratic Majorana Liouville operator Eq.~(\ref{equ:ME}). The late time limit and the steady state here play the role of a low energy limit and the ground state in equilibrium, respectively, where the physics of weakly correlated  superconductors is universally described in terms of Bogoliubov quasiparticle excitations for a variety of microscopic models. Our prescription may thus be seen as a microscopic "parent Liouvillian", providing one possible microscopic realization of a Liouville operator which generates the desired properties at and close to steady state. 

The implementation idea follows closely an earlier proposal for bosons \cite{Diehl08} and is based on a conspiracy of laser driving and engineered dissipation made possible by immersion of the fermionic target system into a superfluid BEC reservoir. The interaction with the bosonic dissipative reservoir is microscopically based on a conventional s-wave fermion-boson density-density interaction and stands in marked contrast to the proximity effect to a BCS superconductor exploited in solid state implementation proposals of the (Hamiltonian) Majorana wire. Our scheme is illustrated in Fig. \ref{fig2} and explained in more detail in the appendix. It yields the following number conserving Liouville dynamics,
\begin{eqnarray}\label{Jump}
\mathcal L[\rho] &=& \tilde\kappa \sum_i [ J_i \rho J_i^\dag - \tfrac{1}{2}\{J_i^\dag J_i,\rho \} ],\,\,
\\\nonumber
J_i &=& \tfrac{1}{4} (a_i^\dag + a_{i+1}^\dag) (a_i-  a_{i+1}).
\end{eqnarray}
These Lindblad operators give rise to \emph{dissipative pairing} in the absence of any conservative forces, a mechanism based on an interplay of phase locking and Pauli blocking established recently \cite{Diehl10b}. The relation to the Majorana operators is apparent in the thermodynamic limit, where it can be shown that the following general relation between fixed number ($J_i$) and fixed phase ($j_i$) Lindblad operators holds (see appendix),
\begin{eqnarray}\label{Equiv}
J_i = C_i^\dag A_i \Leftrightarrow j_i = C_i^\dag + A_i.
\end{eqnarray}
Here $C_i^\dag (A_i)$ are creation (annihilation) parts, respectively, which for Eq. (\ref{Jump}) read $C_i^\dag = \tfrac{1}{2} (a_i^\dag + a_{i+1}^\dag) , A_i = \tfrac{1}{2} ( a_i-  a_{i+1})$. In consequence, we see that indeed precisely Kitaev's quasiparticle operators are obtained as Lindblad operators, $j_i = \tilde a_i$. The description in terms of fixed phase operators becomes appropriate at late times, i.e., close to the steady state, where an ordering principle is provided by the macroscopic occupation of only a few correlation functions. The explicit calculation shows that Eq. (\ref{equ:ME}) is produced with effective dissipative rate $\kappa = \tilde\kappa/8$ (see appendix). There we also discuss the leading imperfections, showing that they preserve the chiral symmetry necessary to remain in the above described topological class. 

\section{Conclusions} 

In this work, we established a complete list of topological features familiar from ground state physics of certain Hamiltonians in engineered dissipative dynamics, highlighting the universality of the concept of topological order in quantum mechanical many-body systems. While the present work has focused on the conceptually simplest system of a quantum wire, the idea of dissipatively induced topological order is more general and will be present in other physical systems and in dimensions higher than one.

\section*{Acknowledgments}
We  thank C. Bardyn, V. Gurarie, A. Imamoglu, C. Kraus and M. Troyer for helpful discussions. We acknowledge support by the Austrian Science Fund (FOQUS), the European Commission (AQUTE, NAMEQUAM), the Institut f\"ur Quanteninformation GmbH, and by a grant from the US Army Research Office with funding from the DARPA OLE program.

\appendix

\section{Appendix}

\subsection{Implementation in cold atomic gases}

\emph{Microscopic Model} -- Here we specify a physical setup leading to Eq.~(\ref{Jump}), following an earlier proposal for bosons \cite{Diehl08} and illustrated in Fig.~\ref{fig2}. We start from an optical superlattice setting, with  lower sites which make up the fermion wire and correspond to the lowest Bloch band of the lattice, and auxiliary sites associated to the second Bloch band located on the links.  These two Bloch bands are coupled via driving lasers, whose Rabi frequencies $\pm \Omega$ are chosen with opposite sign for each pair of lower sites. This amounts to a commensurability condition of driving and lattice laser, and ensures the relative minus sign in the annihilation part of the Lindblad operators of Eq.~(\ref{Jump}), $A_i \propto a_i -a_{i+1}$. By immersing the whole setting into a BEC reservoir, particle superpositions in the upper band can spontaneously decay back to the lower one by emission of a Bogoliubov phonon into the BEC bath. This process is isotropic and short-ranged for suitable bath parameters, giving rise to the two-site creation part in Eq. (\ref{Jump}) with relative plus sign, $C_i \propto a_i + a_{i+1}$. Microscopically, this dissipative mechanism requires a standard s-wave fermion-boson interaction between system (fermions in the optical superlattice) and bath (BEC) particles \cite{Diehl08}. An effective single particle microscopic model such as Eq.~(\ref{Jump}) is obtained by integrating out the upper band under suitable detuning conditions for the driving laser. Using the recently developed tools of single site addressability for optical lattices \cite{Greiner10,Kuhr10}, an edge can be constructed by cutting the superlattice in the right place cf.~Fig.~\ref{fig2}. 

Eq.~(\ref{Jump}) describes an interacting, number conserving Liouville dynamics, which leads to dissipative pairing \cite{Diehl10b}, described by a pure BCS-type wavefunction for paired fermions. There are two ways of seeing this, either by explicit construction of the dark state wavefunction (fixed particle number) or by deriving a suitable mean field theory (fixed phase). In the thermodynamic limit both procedures are equivalent. Here we follow the second option, and highlight the more precise connection between the two approaches below.

\emph{Mean field theory at late times} -- At late times, following Ref.~\cite{Diehl10b} we can make use of the proximity to the steady state to derive a quadratic mean field theory for this dissipative dynamics. In the spirit of BCS theory, the ordering principle behind this approximation is the macroscopic occupation of only a few correlation functions in steady state, which can be evaluated explicitly on the exactly known steady state. To implement the approximation, as usual in BCS type mean field theories, we give up exact particle number conservation and work with a fixed phase. 

We thus start from number conserving Lindblad operators $J_i = C_i^\dag A_i$, where the creation part $C_i^\dag = \sum_j v_{i-j} a_j^\dag$ and the annihilation part $A_i = \sum_j u_{i-j} a_j$ with translation invariant complex position space functions $v_{i-j},u_{i-j}$; in the example Eq.~(\ref{equ:jump}) discussed in the text, $v_{i-j} = \tfrac{1}{2} (\delta_{ij} + \delta_{i+1,j}), u_{i-j} = \tfrac{1}{2} (\delta_{ij} - \delta_{i+1,j})$. For the practical calculation, we switch to momentum space, where we have $J_k = \sum_i e^{\mathrm i kx_i}J_i = \sum_q C^\dag_{q-k} A_q$. The Fourier transforms for the creation and annihilation part are local in momentum space, $C^\dag_k = \sum_i e^{-\mathrm i kx_i} C^\dag_i = v_k a^\dag_k, A_k = \sum_i e^{\mathrm i kx_i} A_i = u_k a_k$, with $v_k = e^{\mathrm i k/2}\cos \tfrac{k}{2}, u_k = \mathrm i e^{-\mathrm i k/2}\sin \tfrac{k}{2}$ in the example (due to the structure of the Liouville operator, the exponential prefactors are irrelevant and can be omitted). In the following we work with functions with the property $u_{-k} = \pm u_k , v_{-k} =\mp v_k$, which however do not necessarily obey a normalization constraint.

For a setting with fixed phase, we now make an ansatz for the steady state density matrix of the product form $\rho = \prod_{k\geq 0}\rho_k$, with density matrix for each mode pair $\rho_k = (u_k +e^{\mathrm{i}\theta}v_{k}a_{-k}^{\dag}a_{k}^{\dag})|\text{vac}\rangle\langle\text{vac}|(u^*_k +e^{-\mathrm{i}\theta}v^*_{k}a_{k}a_{-k})$. Inserting this ansatz into Eq. (\ref{Jump}) in momentum space, and using the projection prescription $\rho_k = \mathrm{tr}_{\neq \pm k}\rho$ on the mode pair $\pm k$, we obtain the equations of motion for the single pair density matrices in the presence of nonzero mean fields. These result from the coupling to other momentum modes, with values determined by the steady state properties in the late time dynamics. 
The resulting mean field dynamics close to the steady state is given by
\begin{eqnarray}\label{Late}
\partial_t \rho &=& \sum_k  \kappa_k\big(j_k \rho j_k^\dag - \tfrac{1}{2}\{j_k^\dag j_k,\rho \}\big) ,\\\nonumber
j_k &=& u_k a_k  + v_k a_{-k}^\dag, \, \kappa_k= \tilde\kappa (|u_k|^2 + |v_k|^2 ) \int \tfrac{dq}{2\pi}|v_q u_q|^2,
\end{eqnarray}
where $j_k$ are fermionic quasiparticle operators up to a normalization, obeying the anti-commutation relations $\{j_k,j^\dag_q \} = (|u_q|^2 + |v_q|^2)\delta_{kq}, \{j_k,j_q \} =  \{j^\dag_k,j^\dag_q \} = 0$. In our example, $|u_q|^2 + |v_q|^2=1$ and $\kappa_k= \tilde\kappa/8$, i.e. the dynamics has constant damping rate for all modes.  In fact, transforming back to position space produces precisely Eq. (\ref{equ:jump}), $j_i = \sum_k e^{- \mathrm i kx_i} j_k = \tfrac{1}{2}( a_i - a_{i+1} +  a_i^\dag +  a_{i+1}^\dag)$.

Since $j_k (u_k +e^{\mathrm{i}\theta}v_{k}a_{-k}^{\dag}a_{k}^{\dag})|\text{vac}\rangle =0$, we see that the above ansatz indeed provides the correct steady state solution. In addition, the equivalence of fixed number and fixed phase wavefunctions can now be justified in our nonequilibrium context \emph{a posteriori} in the thermodynamic limit: the fixed phase BCS state has relative number fluctuations  $\Delta N^2 = \tfrac{\langle \hat N^2 \rangle - \langle \hat N \rangle^2}{\langle \hat N \rangle^2} 
\sim \frac{1}{N}$, where $\hat N$ is the particle number operator and $N$ the number of degrees of freedom.

We emphasize that, remarkably, the late time evolution of the master equation (\ref{Jump}) naturally gives rise to quasilocal \emph{squeezing} of fermions. The mechanism behind this effect is in close analogy to a superconductor: the system here acts as its own reservoir by providing an order parameter, which allows for the appearance of the off-diagonal pair annihilation / creation terms $\sim a_k a_{-k},$ h.c. in Eq. (\ref{Late}).

\emph{General relation of fixed number and fixed phase Lindblad operators} -- Here we show Eq. (\ref{Equiv}). Suppose we are given a fixed phase Liouvillian defined by a set of Lindblad operators of a form 
\begin{eqnarray}
j_i = C^\dag_i + A_i
\end{eqnarray}
using the above conventions, such that the momentum space Lindblad operators read $j_k = u_k a_k + v_k a^\dag_{-k}$. The dark state (with property $j_i |BCS,\theta\rangle = j_k |BCS,\theta\rangle =0$ for all $i$ or $k$) is conveniently formulated in momentum space and reads 
\begin{eqnarray}
|BCS,\theta\rangle &\propto& \exp(e^{\mathrm{i}
\theta}G^\dag  )  |\text{vac}\rangle = \prod_{k}(1+e^{\mathrm{i}\theta}\varphi_{k}a_{-k}^{\dag}a_{k}^{\dag})|\text{vac}\rangle, \nonumber\\
 G^\dag &=& \sum_k \varphi_k a_{-k}^\dag a_k^\dag, \quad \varphi_{k} = \frac{v_k}{u_k}= -\varphi_{-k}.
\end{eqnarray}
Now we want to show that the number conserving version is given by the following expressions for the Lindblad operators and related dark state wavefunction,
\begin{eqnarray}
J_i &=& C_i^\dag A_i, \quad 
|BCS,N\rangle \propto G^{\dag\,N}|\text{vac}\rangle.
\end{eqnarray}
To prove it, we proceed in momentum space. For the normal ordered Lindblad operators $J_k = \sum_q C^\dag_{q-k} A_q$, the dark state property $J_i |BCS,N\rangle = J_k |BCS,N\rangle =0$ for all $i$ or $k$ is equivalent to the vanishing of the following commutator for all $k$,
\begin{eqnarray}
[ J_k,G^\dag ] &=& \sum_q v_{q-k} u_q \varphi_q a^\dag_{q-k}a^\dag_{-q} \\\nonumber
&=& - \sum_q v_{q} u_{q-k} \varphi_{q-k} a^\dag_{q-k}a^\dag_{-q} \stackrel{!}{=} 0.
\end{eqnarray}
This is true if and only if  $\frac{v_q  u_{q-k}}{u_q v_{q-k}} = \frac{\varphi_q}{\varphi_{q-k}}$ for all $k$, i.e. for the wavefunction $\varphi_q = v_q/u_q$ as claimed above. 

Note that working with the functions $|BCS,N\rangle$ for finite $N$ requires an infrared momentum cutoff $q_L \sim 1/L$ such that $N = n L$ ($n$ the density of particles), that is consistent with the thermodynamic limit $N\to \infty, L\to \infty, n= N/L \to \mathrm{const}$.

Thus, for a given real space quadratic master equation, we can immediately construct a number conserving version and vice versa by the above arguments, and indicate the respective fixed number of fixed phase exact dark state wavefunctions.

\emph{Imperfections} -- In the above implementation scheme, the annihilation part $A_i = a_i - a_{i+1}$ is 
well under control since it relies on a locking of lattice and driving laser. The dominant imperfection appears in the creation part: In the case that the phonon wavelength in the BEC bath is not smaller than the lattice spacing (subradiant case), the decay may take place over several lattice sites, such that e.g. $C^\dag_i = \tfrac{1}{2} (a^\dag_i + a^\dag_{i+1} + \epsilon (a^\dag_{i-1} + a^\dag_{i+2})$, giving rise to fixed phase Lindblad operators $j_k = N_k^{-1/2} [  (\cos\tfrac{k}{2} + \epsilon \cos \tfrac{3k}{2}  ) a^{\dagger}_{-k} + i \, \sin \tfrac{k}{2} a_{k}  ]$ with $N_k =  1 +  2 \epsilon    \cos \tfrac{k}{2} \cos \tfrac{3k}{2}  + \epsilon^2 \cos ^2\tfrac{3k}{2}$ in momentum space, which exhibit fermionic anticommutation relations guaranteeing a pure steady state. The resulting vector $\vec n_k$ still lies in a plane for arbitrary $\epsilon$, i.e. the imperfection preserves the chiral symmetry.

\subsection{Properties of the finite size system}

\emph{Equations of motion in the Majorana basis} -- For the practical treatment of the finite size system quadratic in the fermion operators and with Gaussian initial conditions, it is convenient to encode the information in the covariance matrix of second moments in the real Majorana basis. For convenience, we repeat and extend here the definitions of the main text. The Majorana operators obey $c_{2j}= (a^\dag_{j}+a_{j})$, $c_{2j-1}= \mathrm i (a^\dag_{j}- a_{j})$, $c^\dag_j=c_j$, $\{c_i,c_j\}=2\delta_{ij}$, where $j=1,... N$ labeling the physical sites in a one dimensional lattice. Following \cite{Eisert11}, we start with a quadratic master equation written in the Majorana basis, i.e. $\partial_t \rho = -i [\mathcal H,\rho] +  \kappa \sum_i j_i \rho j_i^\dag - \tfrac{1}{2} \{j^\dag_i j_i , \rho\}$ with $\mathcal H=\tfrac{1}{4}c^T H c$ for $2N\times 2N$ hermitian matrix $H$ and $j_i =l_{i}^Tc, j^\dag _i = c^T l_{i}^*$, for $2N$ component column vectors $l_i,c$. The Liouvillian parameters are then encoded in a hermitian $2N\times 2N$ matrix $M = \sum_i  l_{i}\otimes l_i^\dag $. The equation of motion for the covariance matrix with entries $\Gamma_{ab} = \tfrac{\mathrm i}{2} \langle [c_a , c_b]\rangle $ reads ($\kappa=1$)
\begin{eqnarray}
\partial_t \Gamma  &=& 
= -   \mathrm i [H, \Gamma] - \{X, \Gamma\}  - Y ,
\end{eqnarray}
with real matrices $X = 2 \mathrm{Re} M = X^T$, $Y = 4 \mathrm{Im} M = -Y^T$. The spectrum of $X$ is positive semidefinite \cite{Prosen08} and a pure state is signaled by the eigenvalues of $\Gamma^2$ being all equal to $-1$.

\emph{Zero modes} -- The real symmetric matrix $X$ is diagonalizable and has real eigenvalues and -vectors. First we show the existence of zero eigenvalues for this matrix for a wide class of finite systems via explicit construction of the corresponding eigenvectors. The ideal Lindblad operators for the dissipative topological 1D quantum wire are characterized by the following vector in the Majorana basis:
\be
l^T_j =  \begin{pmatrix} 0,  & \cdots  , & 0, & (l_j)_{2j}, & (l_j)_{2j+1}, & 0, & \cdots , &  0  \end{pmatrix},
\ee
with $(l_j)_{2j} = \mathrm i$ and $( l_j)_{2j+1} = 1$. Let us now consider the more general vector $ l_j$ specified with
\be
\begin{split}
(l_j)_{2j-1} &= \epsilon_{2j-1}, ~\,~ (l_j)_{2j} =\mathrm  i (1+\epsilon_{2j}), \\
(l_j)_{2j+1} &= (1 + \epsilon_{2j+1}), ~ \, ~ (l_j)_{2j+2} =\mathrm  i \epsilon_{2j+2}.
\end{split}
\ee
For each $l_j$, there are two orthogonal vectors given by 
\be
\begin{split}
\tilde v^T_{L,j} &= \begin{pmatrix} 0, & \cdots , & 1, & 0 ,& - \frac{\epsilon_{2j-1}}{1 + \epsilon_{2j+1}}, & 0, & \cdots  ,& 0 \end{pmatrix}, \\
\tilde v^T_{R,j} &= \begin{pmatrix} 0, & \cdots , & 0 ,& - \frac{\epsilon_{2j+2}}{1 + \epsilon_{2j}} ,& 0 ,&1 , & \cdots , & 0  \end{pmatrix},
\end{split}
\ee
and by construction, they fulfill $l^T_j \cdot  \tilde v_{L,j} = l_j^\dag \cdot \tilde v_{L,j} = l_j^T \cdot \tilde v_{R,j} = l_j^\dag \cdot \tilde v_{R,j}=0$. Due to the structure of the matrix $M$ as a sum of outer products of the $l_j$ vectors, we find precisely two zero modes for any finite system size $N$ given by the $2N$ component vectors 
\be
\begin{split}
\gamma_L& = \sum_{j=1}^{N} v_{L,j} c_{2j-1} , \\
 &  v_{L,j \neq 1} =  \prod_{i=1}^j \left( - \frac{\epsilon_{2i-1}}{1 + \epsilon_{2i+1}} \right), ~ \, ~ v_{L,j = 1} = 1, \\
\gamma_R& = \sum_{j=1}^{N} v_{R,j} c_{2j} , \\
 &  v_{R,j \neq 2N} = \prod_{i=1}^j    \left(   - \frac{\epsilon_{2(N-i)}}{1 + \epsilon_{2(N-i-1)}}  \right) , ~ \, ~ v_{R,j = 2N} = 1,
\end{split}
\ee
where $\gamma_{L, R }$ represent the left / right Majorana modes. The reason behind these two vectors being zero vectors of the matrices $M$ and $M^*$ is that they are orthogonal to the $(2N-2)$ linearly independent vectors $l_j$ and $l^*_j$ in a space of dimension $2N$. Hence, $v_L$, $v_R$, $\{l_j\}$ and $\{l^*_j \}$ form a complete basis in this space.

Let us now focus on the deformations studied in the main text. Here, the Majorana modes are characterized by vectors 

(i) Canonical deformation:
\be
v_{L,2j-1} =  \epsilon^{j-1} ,  ~ \, ~  v_{R,  2j } = \epsilon^{N - j} , 
\ee

(ii) Non-canonical deformation: 
\be
v_{L,2j-1} =  (-\epsilon)^{j-1} ,  ~ \, ~  v_{R,  2j } = \epsilon^{N - j} , 
\ee
with $\epsilon =\frac{ \sin{\theta} - \cos{\theta}}{\cos{\theta}+ \sin{\theta}} $. The associated localization length is then given by $l_\text{loc} = -a (\log |{ \epsilon}|)^{-1}$, with $a$ the lattice constant.

In the case of a deformation being non-homogenous but with value changing randomly from site to site, the zero modes are still present, but the expression for $\gamma_L$ and $\gamma_R$ is given by
\be
v_{L,2j-1} =  \prod_{i=1}^j \epsilon_i   ,  ~ \, ~  v_{R,  2(N-j+1) } = \prod_{i=1}^j \epsilon_{N-j+1}.
\ee

\emph{Dissipative isolation of the subspace} -- As argued in the main text, in addition to the existence of a zero mode subspace, the dissipative evolution must ensure its isolation from the bulk. The conditions for such a situation are readily formulated generally, without reference to the quadratic setting.  For a set of Lindblad operators $J_i$ making up the total Liouvillian $\mathcal L [\rho] = \sum_i J_{i} \rho J_i^\dag  - \tfrac{1}{2} \{ J_i^\dag J_i, \rho\}$, we may introduce projectors on the edge (zero mode) and bulk subspaces, $p$ and $q =1-p$, respectively. A decoupled edge subspace appears if the Lindblad operators $J_i$ are block diagonal, $J_{i,pq}=J_{i,qp} =0$, with the edge block identical to zero, $J_{i,pp}=0$. We then obtain a dissipative evolution for the density matrix in this projection,
\begin{eqnarray}\label{GenEq}
\hspace{-0.4cm} \partial_t \left(   \hspace{-0.1cm}\begin{array}{cc}
 \rho_{pp} & \rho_{pq} \\
 \rho_{qp} & \rho_{qq} 
 \end{array}  \hspace{-0.1cm}
 \right)    \hspace{-0.05cm}=   \hspace{-0.05cm}\sum_j    \hspace{-0.05cm}
\left( \begin{array}{cc}
0  &  \hspace{-0.5cm}- \tfrac{1}{2} \rho_{pq} J^\dag_{j, qq} J_{\ell, qq} \\
- \tfrac{1}{2} J^\dag_{j, qq} J_{j, qq} \rho_{qp} &    \mathcal L_{j, qq}[\rho_{qq}]
 \end{array}
 \right)\hspace{-.1cm}.\nonumber\\
 \end{eqnarray}
The bulk dissipative evolution $\mathcal L_{j, qq} [\rho_{qq}]  = J_{j, qq} \rho_{qq} J^\dag_{j, qq}  - \tfrac{1}{2} \{ J^\dag_{j, qq} J_{j, qq} , \rho_{qq}\}$ has Lindblad form. The density matrix in the $p$ subspace $\rho_{pp}$ is a constant of motion. The coupling density matrix elements $\rho_{qp},$ h.c. damp out according to $\rho_{qp} = e^{- \sum_j J_{j,qq}^\dag J_{j,qq} t} \rho_{qp} (t=0)$, i.e. exponentially fast in the presence of a dissipative gap. 

This situation is indeed present in our quadratic setting. We switch to the spectral representation $X = \sum_r \lambda_r |r \rangle \langle r|$, where $\lambda_r\neq 0$ are the nonvanishing eigenvalues and $|r \rangle$ the corresponding eigenvectors. In addition, we have zero modes $\lambda_\alpha=0$ which do not contribute to the spectral decomposition, with eigenvectors $|\alpha \rangle$. Without loss of generality we choose orthonormal eigenvectors $\langle a  | b\rangle = \delta_{ab}$. In this basis, the equations of motion read
\begin{eqnarray}\label{specevol}
\partial_t \left(\hspace{-0.1cm}\begin{array}{cc}
\Gamma_{\alpha\beta}  & \Gamma_{\alpha s}\\
\Gamma_{r\beta} & \Gamma_{rs} 
\end{array} \hspace{-0.1cm}\right)\hspace{-0.1cm}
&=&\hspace{-0.1cm} \left( \hspace{-0.1cm}\begin{array}{cc}
0 & - \lambda_s \Gamma_{\alpha s}   \\
- \lambda_r \Gamma_{r\beta}  & - (\lambda_r + \lambda_s) \Gamma_{rs} -  Y_{rs} 
\end{array}\hspace{-0.1cm} \right),\nonumber\\
\end{eqnarray}
where $Y_{ab} = \langle a | Y | b\rangle$. Following the above discussion, the zero modes of the matrix $X$ are also zero modes of the matrix $Y$ and vice versa. This implies $Y_{\alpha s}= Y_{ r\beta } =0$ and shows the decoupling of the edge and bulk subspaces, as well as $Y_{\alpha\beta}=0$, which defines the zero modes subspace. This reflects the structure of Eq. (\ref{GenEq}) for a quadratic theory in the Majorana basis.

\emph{Adiabatic parameter changes} -- We study the evolution for a time dependent Liouville (or Hamilton) operator in the instantaneous basis in the quadratic setting, where the form of (\ref{specevol}) is kept but the eigenvalues and -vectors of $X$ now depend on time. The transformation into the instantaneous basis $|a(t)\rangle = U(t) |a_0\rangle $, where $|a_0\rangle$ is the initial reference basis, is now orthogonal. 
The connection is given by the matrix $\bar A \equiv \langle a(t) | \dot b(t) \rangle =  \langle a_0 | U^{\dagger}\dot U |b_0 \rangle$, which is antisymmetric due to normalization, $\partial_t \langle a  |b \rangle=0$. With Hamiltonian in the instantaneous basis given by $h_{ab} =   \langle a  | H| b \rangle$ and hermitian matrix $A = \mathrm i \bar A$, the equation of motion in this basis reads
\begin{eqnarray}
\partial_t \Gamma 
= - \mathrm i [  h + A,\Gamma ] - \{ \lambda,  \Gamma\}     - Y.
\end{eqnarray} 
As already mentioned in the main text, the shift in the Hamiltonian $h \to h +A$ shows the emergence of a gauge structure due to the explicit time dependence of the eigenbasis, irrespective to whether the time evolution of $\Gamma$ is generated by a Hamiltonian or Liouvillian.

We study the simple example of local adiabatic parameter changes from the main text: First, we consider a system of two physical sites, specified by the vector $l_\ell  = \tfrac{1}{2} (0, \cos{\theta},\mathrm i, - \sin{\theta})$. The spectrum of $X$ is then doubly degenerate with eigenvalues $0,1/2$. The connection matrix is $\bar{A} = \dot \theta(1-\sigma_z)\otimes \sigma_y$. The resulting equations of motion for the elements with nonzero rhs are,
\begin{eqnarray}
\partial_t \Gamma_{12} &=& -\dot \theta \Gamma_{14}, \quad \partial_t \Gamma_{14} = -\frac{\Gamma_{14}}{2} + \dot \theta \Gamma_{12} , \\\nonumber
 \partial_t \Gamma_{23} &=& -\frac{\Gamma_{23}}{2} + \dot \theta \Gamma_{34} , \quad \partial_t \Gamma_{34} = -\Gamma_{34} -  \dot \theta \Gamma_{23} -1 .
\end{eqnarray}
The system is easily solved using adiabatic elimination in the limit where the parameter changes are adiabatic, i.e. $\dot \theta(t)\ll1$ for all times. In particular, we then obtain for the slowly evolving variable in the zero eigenvalue subspace 
\begin{eqnarray}
\partial_t \Gamma_{12} &=&  -2 \dot \theta^2  \Gamma_{12} \end{eqnarray}
with solution (written with dimensions restored)
\begin{eqnarray}\label{2siteex}
\Gamma_{12} (t) = \Gamma_{12} (0) \exp \left(-2 \kappa^{-1} \int_0^T dt'  \dot \theta^2 \right),
\end{eqnarray}
describing a dephasing of the Majorana mode. Actually, for linear time changes $\theta(t) = \alpha t$ the integral approaches $\alpha^2 T/\kappa  \ll 1$, i.e. the Majorana mode population is only weakly affected. 

Now, we observe that this result also applies to the general case of $N$ sites. This is due to the structure of the steady state resulting from the ideal case described by Eq.~(\ref{equ:jump}) with Majorana modes paired between different physical sites. The only time dependent changes occur on the Majorana sites $1, 2N-2, 2N-1 , 2N$, while the rest of the system remains decoupled (the situation is visualized e.g.~in Fig.~\ref{Fig1_Setup}). Therefore, the result (\ref{2siteex}) is general.

We thus show that the Majorana subspace remains protected for adiabatic parameter changes also in the dissipative case. The argumentation will be valid and useful in any situation where there is a gap in the damping spectrum.

\section{Topological Invariant and Master Equation}

Here we present some details on the master equation and topological
invariant for a translation invariant quadratic Liouville operator with
chiral symmetry in terms of the density matrix.

\emph{Density Matrix} -- We consider a translation invariant setting, where
this property applies to both the initial state and the Lindblad operators
defining the Liouvillian. For a quadratic Liouvillian, the density matrix
then takes a product form in momentum space $\rho =\prod_{k\geq 0}\rho%
_{k}$, where $\rho_{k}$ is the density matrix of the momentum mode pair $\pm k$%
. Using fermion superselection rules, the $4\times 4$ matrix $\rho_{k}$
can be written in a block-diagonal form%
\begin{equation}
\rho_{k}=%
\begin{pmatrix}
\rho _{1k} & 0 \\ 
0 & \rho _{2k}%
\end{pmatrix}%
,  \label{general_rho}
\end{equation}%
where $\rho _{1k}=\mathrm{diag}(\rho _{+k}^{(1)},\rho _{-k}^{(1)})$ is a
diagonal $2\times 2$ matrix with non-negative elements $\rho _{\pm k}^{(1)}$
in the subspace with odd total occupation of the modes $+k$ and $-k$ equal $1$, $%
\langle a_{k}^\dag a_k\rangle +\langle a_{-k}^\dag a_{-k}\rangle =1$, and $\rho _{2k}$ is a $2\times 2$ hermitian matrix in the even occupation 
subspace $\langle a_{k}^\dag a_k\rangle +\langle a_{-k}^\dag a_{-k}\rangle=0,2$. Taking normalization ($\rho _{+k}^{(1)}+\rho
_{-k}^{(1)}+\mathrm{tr}\rho _{2k}=1$) and hermiticity into account, the
matrix $\rho _{2k}$ can be written as 
\begin{equation}
\rho _{2k}=\mathrm{tr}\rho _{2k} \tfrac{1}{2}(\mathbf{1}+\widetilde{Q}%
_{k}) ,  \label{rho_2k}
\end{equation}%
where $\widetilde{Q}_{k}$ is a traceless hermitian matrix, $\widetilde{Q}%
_{k}=\vec{n}_{k}\vec{\sigma}$ with $\vec{\sigma}$ being the vector of Pauli
matrices and $\vec{n}_{k}$ a real 3-component vector $0\leq \left\vert \vec{n%
}_{k}\right\vert \leq 1$. For a pure state one has $|\vec{n}_{k}|=1$ for all 
$k$. In this case $\rho _{\pm k}=0$, $\mathrm{tr}\rho _{2k}=1$, and $\rho
_{2k}$ takes the form of a projector, $\rho _{2k}^{2}=\rho _{2k}$. The
situation $\vec{n}_{k}=0$ for all $k$ signals a completely mixed state.

It is convenient to introduce the $2\times 2$ matrix $\tilde \rho _{2k}=(\mathrm{tr}%
\rho _{2k})^{-1}\rho _{2k}$, such that 
\begin{equation}
\tilde\rho _{2k}=\tfrac{1}{2}(\mathbf{1}+\widetilde{Q}_{k}),
\end{equation}%
which carries all topological information encoded in the density matrix $%
\rho_{k}$. This is because the other quantities entering $\rho_{k}$,
namely $\rho _{+k}^{(1)},\rho _{-k}^{(1)}$, and $\mathrm{tr}\rho _{2k}$, are
just non-negative numbers subject to the constraint  $\rho
_{+k}^{(1)}+\rho _{-k}^{(1)}+\mathrm{tr}\rho _{2k}=1$, and the space defined
by these conditions is topologically trivial.

The matrix $\rho _{k}$ and, therefore, the vector $\vec{n}_{k}$
are defined for $k\geq 0$. It is convenient, however, to extend them also to
negative values of $k$. This can easily be done by noticing that under the
change $k\rightarrow -k$ the elements $\left\vert 0_{k},0_{-k}\right\rangle $
and $\left\vert 1_{k},1_{-k}\right\rangle $ of the basis in the even occupation subspace 
change as $\left\vert 0_{k},0_{-k}\right\rangle
\rightarrow \left\vert 0_{-k},0_{k}\right\rangle =\left\vert
0_{k},0_{-k}\right\rangle $ and $\left\vert 1_{k},1_{-k}\right\rangle
\rightarrow \left\vert 1_{-k},1_{k}\right\rangle =-\left\vert
1_{k},1_{-k}\right\rangle $. As a result, the off-diagonal elements of $%
\widetilde{Q}_{k}$ change their sign, while the diagonal ones remain
unchanged. That is, we have the transformations 
\begin{equation}\label{trafs}
\widetilde{Q}_{-k}=\sigma _{z}\widetilde{Q}_{k}\sigma _{z},\quad \vec{n}%
_{-k}=S_{z}\vec{n}_{k},
\end{equation}%
with $S_{z}=\mathrm{diag}(-1,-1,1)$. Note that the thus defined vector  $\vec{n}%
_{k}$ is continuous at $k=0$ and $k=\pm \pi $ because $\widetilde{Q}_{k=0}$
and $\widetilde{Q}_{k=\pm \pi }$ have only diagonal elements ($\vec{n}_{k}$
has only $z$-component) due to the fermionic nature of the particles. If in
addition  $\left\vert \vec{n}_{k}\right\vert \neq 0$ for all $k$, the vector 
$\vec{n}_{k}$ defines a mapping of the Brillouin zone (topologically
equivalent to $S^{1}$) into a sphere $S^{2}$, $k\rightarrow \hat{\vec{n}}%
_{k}=\vec{n}_{k}/\left\vert \vec{n}_{k}\right\vert $.

\emph{Topological Invariant} -- The mappings constructed above are topologically trivial because the
homotopy group $\pi _{1}$ of a sphere $S^{2}$ is trivial, $\pi _{1}(S^{2})=0$%
. In other words, every mapping $S^{1}\rightarrow S^{2}$ can be continuously
deformed into a trivial one that maps the entire Brillouin zone into a point
on a sphere ($\hat{\vec{n}}_{k}=\hat{\vec{n}}$ is $k$-independent).
As a result, a general density matrix of the form (\ref{general_rho}) does not
have topological order. The situation is different for the density matrices
obeying the chiral symmetry \cite{Altland97,Ryu10}. In this case, there
exists a unitary matrix $\Sigma $ that anticommutes with  $\widetilde{Q}_{k}$
for all $k$,%
\begin{equation}
\Sigma \widetilde{Q}_{k}+\widetilde{Q}_{k}\Sigma =0,\quad \Sigma ^{2}=\mathbf{1}.
\label{chiral_symmetry}
\end{equation}%
The condition of the chiral symmetry in our case has a simple geometrical
interpretation. Namely, if we write the matrix $\Sigma $ in the form $\Sigma
=\vec{a}\vec{\sigma}$, where $\vec{a}$ is a real unit vector, $\left\vert 
\vec{a}\right\vert =1$, then Eq. (\ref{chiral_symmetry}) is equivalent to%
\begin{equation}
\vec{a}\vec{n}_{k}=0,
\end{equation}%
i.e. the vector $\vec{n}_{k}$ for all $k$ belongs to the plane that is
orthogonal to the vector $\vec{a}$. As a result, the vector $\hat{\vec{n}}%
_{k}$ determines a mapping of the Brillouin zone ($S^{1}$) into a circle $%
S^{1}$, which is an interception of the unit sphere $S^{2}$ with the plane
passing through its origin (one of the great circles). The mappings of this
type can be divided into topologically different classes that are
characterized by an integer winding number (\ref{winding}) or (\ref{windingQ}%
) because the homotopy group in this case is nontrivial, $\pi _{1}(S^{1})=%
\mathbf{Z}$.  

To construct the topological invariant that distinguishes topologically
different density matrices we use the fact that $\widetilde{Q}_{k}^{2}=\vec{n%
}_{k}^{2}\mathbf{1}$ and introduce the normalized matrix $Q_{k}=\left\vert 
\vec{n}_{k}\right\vert ^{-1}\widetilde{Q}_{k}$ with a unit square, $%
Q_{k}^{2}=\mathbf{1}$. This matrix can always be written as%
\begin{equation}
Q_{k}=U^{\dagger }%
\begin{pmatrix}
0 & e^{-\mathrm i\phi _{k}} \\ 
e^{\mathrm i\phi _{k}} & 0%
\end{pmatrix}%
U
\end{equation}%
or%
\begin{equation}
Q_{k}=V^{\dagger }%
\begin{pmatrix}
m_{z,k} & m_{x,k} \\ 
m_{x,k} & -m_{z,k}%
\end{pmatrix}%
V,
\end{equation}%
where $\phi _{k}$, $m_{x,k}$, and $m_{z,k}$ are real, $%
m_{x,k}^{2}+m_{z,k}^{2}=1$, with some unitary matrices $U$ and $V$ (in the
first case $\Sigma =\pm U^{\dagger }\sigma _{z}U$, in the second case $%
\Sigma =\pm V^{\dagger }\sigma _{y}V$). Therefore, all information about the
stationary state is encoded in the phases $\phi _{k}$ or in the unit vectors 
$\vec{m}_{k}=(m_{x,k},0,m_{z,k})$ in the ($x-z$)-plane. As a result, the
topologically different classes of stationary states with the chiral
symmetry can be classified by an integer winding number%
\begin{eqnarray}
W[\rho ] &=&\frac{1}{\pi }\int_{0}^{\pi }d\phi _{k}=\frac{\phi _{\pi }-\phi
_{0}}{\pi }  \label{windingVec} \\
&=&\frac{1}{\pi }\int_{0}^{\pi }dk(m_{z}\frac{dm_{x}}{dk}-m_{x}\frac{dm_{z}}{%
dk})  \notag \\
&=&\int_{\mathrm{BZ}}\frac{dk}{2\pi }(m_{z}\frac{dm_{x}}{dk}-m_{x}\frac{%
dm_{z}}{dk}),  \notag
\end{eqnarray}%
where in the last line we extend the integration over the entire Brillouin
zone $k\in \lbrack -\pi ,\pi ]$ using the continuation from positive to
negative $k$ as discussed above, cf. Eq. (\ref{trafs}). We see that $%
W[\rho ]$ indeed indicates the number of times the end of the vector $\vec{m}%
_{k}$ winds around the origin when $k$ goes over the Brillouin zone $k\in
\lbrack -\pi ,\pi ]$.

In terms of the matrix $Q_{k}$,$\,$the invariant can be written as%
\begin{equation}
W[\rho ]=\frac{1}{4\pi \mathrm{i} }\int_{\mathrm{BZ}}dk\,\mathrm{tr}\left(
\Sigma Q_{k}\partial _{k}Q_{k}\right) .  \label{windingQ}
\end{equation}%
Inserting $\Sigma =\vec{a}\vec{\sigma},Q_{k}=\hat{\vec{n}}_{k}\vec{\sigma}$
into Eq. (\ref{windingQ}) yields Eq. (\ref{winding}) in the main text.

Note that the definition of the topological invariant depends on the vector
$\vec{a}$, which can only be defined up a sign ($\Sigma$ and $-\Sigma$ are
equivalent). While $k$ independent, this vector can in general depend on the parameters of the
Liouvillian. As a result, the global definition of $\nu$ requires the existence of a continuous choice of one of the two branches of $\vec{a}$ in the entire parameter space. Whether it is possible or not, depends on the
parameter space itself: One has to be able to connect any two points by some
deformation path, along which one can define $\vec{a}$ continuously. If this
is not the case, the global choice of $\vec{a}$ and, hence, the global
definition of $\nu$ is not possible, and one has arbitrarily chosen $\vec{a}$ in
different "disconnected" parts of the parameter space. The relative sign of
$\nu$ in different parts has then no significance. In any case, however, the
nonzero value of $\nu$ indicates nontrivial topological order, irrespective to its sign.

\emph{Equation of Motion and Solution} --  Using the product form of the density matrix, applying the projection $\rho_k = \mathrm{tr}_{\neq \pm k}\rho$ (all mode pairs but $\pm k$ are traced over) we obtain the equation of motion for each pair, which reads ($\kappa=1$)
\begin{eqnarray}\label{LiouvillianGamma}
\partial_t \rho_k &=&  \sum_{\sigma = \pm} j_{\sigma k} \rho_k j_{\sigma k}^\dag - \tfrac{1}{2} \{j_{\sigma k}^\dag j_{\sigma k} , \rho_k\} 
,\\\nonumber
j_k &=&  u_k c_k + v_k c_{-k}^\dag = \xi_k^T \Psi_k, \, \xi_k =\hspace{-0.1cm}
\left(\begin{array}{c}
u_k\\
v_k
\end{array}\right)
, \, \Psi_k\hspace{-0.1cm} = \hspace{-0.1cm}\left(\begin{array}{c}
c_k\\
c_{-k}^\dag
\end{array}\right)
\end{eqnarray}
for $k\neq 0,\pm\pi$. Here $u_k,v_k$ are arbitrary complex functions of momentum; in particular, they need not to fulfill  a normalization condition. For the solution, we study the set of single particle correlation functions (covariance matrix) which is closed due to the quadratic nature of (\ref{LiouvillianGamma}). Using suitable symmetrizations, the equation of motion for the correlation functions can be written in terms of four component vectors
\begin{eqnarray}\label{TransinvEvol}
\hspace{-0.4cm} \partial_t \vec N_k &=&  - \kappa_k ((\mathbf{1} + A_k) \vec N_k - \vec M_k^s ), \,
A_k = 
\left(\begin{array}{cc}
0 & \vec m_k^{a\,T} \\
\vec m_k^a & 0  
\end{array}\right)\hspace{-0.1cm},\nonumber\\
\end{eqnarray} 
where the vector $\vec{N}_k$ is defined in terms of the single particle correlation functions, and we use the further definitions
\begin{widetext}
\begin{eqnarray}
\vec N_k &=&  \left(\begin{array}{c}
n_{0,k}  \\
\vec n_k
\end{array}\right) =
\left(\begin{array}{c}
\tfrac{1}{2} (C_k - C_{-k}) \\
\langle c_{-k}^\dag c_k^\dag\rangle +\langle c_{k} c_{-k}\rangle \\
\mathrm i (\langle c_{-k}^\dag c_k^\dag\rangle -\langle c_{k} c_{-k}\rangle ) \\
\tfrac{1}{2} (C_k + C_{-k})
\end{array}\right), \quad C_k = \langle c_k c_k^\dag\rangle  - \langle c_{k}^\dag c_k\rangle,
\\\nonumber
m_{\mu,k} &=& \frac{\tilde m_{\mu,k}}{\kappa_k}, \quad \tilde m_{\mu,k} = \xi^\dag_k \sigma^\mu \xi_k, \quad \sigma^\mu = (\mathbf{1}, \sigma^i),\quad \mu =0, ... , 3 ,\quad 
\kappa_k = \tfrac{1}{2} (\tilde m_{0,k} + \tilde m_{0,-k}), \quad\\\nonumber 
\vec M^s_k &=& 
\left(\begin{array}{c}
m^s_{0,k} \\
\vec m_k^s 
\end{array}\right) = 
\frac{1}{2}\left(\begin{array}{c}
m_{0,k} - m_{0,-k} \\
m_{x,k} -m_{x,-k} \\
m_{y,k} -m_{y,-k}\\
m_{z,k}  + m_{z,-k}
\end{array}\right),\quad 
\vec M^a_k = 
\left(\begin{array}{c}
m^a_{0,k} \\
\vec m^a_{k} 
\end{array}\right) = 
\frac{1}{2}\left(\begin{array}{c}
0 \\
m_{x,k}  + m_{x,-k} \\
m_{y,k} + m_{y,-k}\\
m_{z,k}  - m_{z,-k}
\end{array}\right),
\end{eqnarray}\label{MESol}
\end{widetext}
and we used $\langle 1 \rangle =1$. 
We summarize some properties of these equations: \\
(i) By construction, all components of $\vec N_k, \vec M^{a/s}_k$ are even or odd eigenfunctions under momentum reflection. In particular, 
\begin{eqnarray}
\vec n_{-k} = S_z\vec n_{k}, \, \vec m^s_{-k} = S_z\vec m^s_{k}, \, \vec m^a_{-k} = - S_z\vec m^a_{k}.
\end{eqnarray}
(ii)  Despite the appearance of four components in Eq. (\ref{TransinvEvol}),  all physical information on the steady state can be stored in the vector $\vec n_k$. This is because given this knowledge, the first line in this equation only carries redundant information: $n_{0,k} = m^s_{0,k}   - \vec n^T_k \vec m_k^{a}$. The general time dependent solution may need four real numbers for each mode pair $\pm k$ for a complete description.\\
(iii) \emph{Fermionic operators} -- Important differences between the steady states occur depending on whether the Lindblad operators describe fermionic quasiparticle operators (up to a normalization). More precisely, the anticommutation relations are 
\begin{eqnarray}\label{quasicanonical}
\{ j_k , j_q^\dag \} &=& \xi_k^\dag \xi_k \,\delta_{k,q}, \,\\\nonumber
 \{ j_k , j_q \} &=& (u_k v_{-k} + u_{-k} v_{k} )\, \delta_{k,-q}, \,\, \text{h. c. }
\end{eqnarray}
Introducing $\tilde j_k = j_k/(\xi_k^\dag \xi_k)^{1/2}$, the first anticommutator can be normalized if $\xi_k^\dag \xi_k$ is strictly positive. The Dirac algebra is then fulfilled if the equal charge operators anticommute, i.e. for 
\begin{eqnarray}
u_k v_{-k} = - u_{-k} v_k.
\end{eqnarray}
If these conditions are fulfilled, we call the deformation from the ideal Lindblad operators Eq. (\ref{equ:jump}) \emph{quasi-canonical}, as it is related to a true canonical transformation by a simple rescaling. (In the specific quasi-canonical deformation considered in the main text, Eq. (\ref{deformations}), even the stronger conditions $u_{-k} =- u_k,  v_{-k} =  v_k$ apply.) In this case, the solution simplifies since the components $m_{\mu, k}$ themselves are odd or even eigenfunctions under momentum reflection, notably $ \vec m_{-k} = S_z  \vec m_k$. We then have $\vec M^{s\,T}_k  = (0, \vec m_k^T ), \vec M^{a\,T}_k = (0,\vec 0^T)$. In this case, the steady state solution simplifies to $\vec n_k = \vec m_k = \xi^\dag_k \vec \sigma \xi_k/m_{0,k}$ fulfilling the purity condition $|\vec n_k| =1$ for all $k$, such that the density matrix takes the form of a projector, $\rho_k^2=\rho_k$. This density matrix then describes the ground state of a Hamiltonian $H = \sum_k H_k, H_k = \tfrac{1}{2} \Psi_k^\dag \vec m_k \vec \sigma\Psi_k$. In contrast, when the vector $\vec m_k$ encoding the structure of the Liouvillian does not have the above transformation behavior, $\vec m_{-k} \neq S_z  \vec m_k$ for some modes, the steady state is mixed. \\
(iv) \emph{Current-free situation} -- For $n_{0,k}=0$ the system carries not current, $\langle \hat J_{k=0}\rangle =0$, where the current operator in momentum space reads $\hat J_k = e^{-\mathrm i k/2}\sum_q \sin q \,a_{q-k/2}^\dag a_{q+k/2}$. Solutions of this type are obtained for the cases $\tilde{\vec M}_{-k}^2 = \tilde{\vec M}_{k}^2$ or $\tilde m_{0,-k} = \tilde m_{0,k}$,  and $\tilde{\vec M}_{-k}\cdot \tilde{\vec M}_k = 2 \tilde m_{z,-k}\tilde m_{z,k}$. The first condition has a simple interpretation in terms of symmetric damping of $\pm k$ modes, similar to the symmetry in the excitation spectrum of a Hamiltonian in the absence of a magnetic field.  In this case, the steady state solution simplifies to $\vec n_k = \vec m^s_k, m_{0,k}^s=0$, and in general describes a mixed state $|\vec n_k|\leq 1$, which cannot be written as a ground state of a Hamiltonian. We note that for fermionic $j_k$ (up to normalization), this condition is always fulfilled and the system is current-free. \\
(v) The solution of the linear equation is given by 
\begin{eqnarray}
\vec N_k(t) = e^{- L_kt} \vec N_k(0) + (\mathbf{1} - e^{-\kappa_k L_k t} )L_k^{-1} \vec M_k
\end{eqnarray}
with $L_k =\kappa_k(\mathbf{1} + A_k)$. Clearly, symmetries of the time evolving density matrix are determined both by the initial state $ \vec N_k(0)$ and the Liouville operator parameters, i.e. $L_k, \vec M_k$. For a unique steady state (all eigenvalues of $L_k$ strictly positive), all memory of the initial state is lost and thus the symmetry of the Liouvillian is inherited by the steady state density matrix. If a symmetry is shared by both initial state and Liouville operator, that property is conserved during the evolution. In particular, we note that generic initial states like Gaussian thermal states are chirally symmetric.\\
(vi) The damping spectrum for the correlation functions is determined by the eigenvalues of $\kappa_k (\mathbf{1} + A_k)$. In the case of the Lindblad operators being fermionic quasiparticle operators, it is fourfold degenerate and coincides with $\kappa_k$. More generally, the spectrum is given by
\begin{eqnarray}\label{spectras}
\lambda^{1,2}_k = \kappa_k, \quad\lambda^{3,4}_k = \kappa_k (1 \pm | \vec m_k^a|),
\end{eqnarray}
with the first eigenvalue doubly degenerate. Positive semidefiniteness is seen from $| \vec m_k^a|\leq 1$.

\emph{Examples} -- In the main text we discuss two examples with quasilocal (nearest neighbour) Lindblad operators, the quasi-canonical and the non-canonical deformation. In momentum space, their extended versions are given by $j^{(\ell)}_k = \xi_k^{(\ell)}\Psi_{k}$ with $\ell = c,n$ for the canonical and non-canonical deformation,
\begin{eqnarray}
\xi_k^{(c)} (\theta,\phi)&=& \sqrt{2}\left(\begin{array}{c} 
- i e^{i \phi}  \sin{\theta} \, \sin \tfrac{k}{2}  \\
  \cos{\theta} \,  \cos \tfrac{k}{2}
\end{array}\right) , \\\nonumber
\xi_k^{(n)} (\theta,\phi) &=&\tfrac{1}{\sqrt{2}} \left(\begin{array}{c} 
 e^{i \phi} ( \cos{\theta}  e^{- \mathrm i k/2} - \sin \theta e^{ \mathrm i k/2 } )\\ 
\cos{\theta}  e^{\mathrm i k/2} + \sin \theta e^{- \mathrm i k/2 }
\end{array}\right) ,
\end{eqnarray}
steady state solutions 
\begin{eqnarray}
\vec{n}_k^{(c)} &=&\vec m_k =  \frac{ 1}{1+ \cos(2 \theta) \cos{k} }\begin{pmatrix}
\sin\phi \sin{(2 \theta)} \sin{k}   \\
 \cos\phi \sin{(2 \theta)} \sin{k}\\
-(\cos{(2 \theta)} + \cos{k})
\end{pmatrix},\nonumber\\
\vec{n}_k^{(n)}  &=& \vec m^s_k = \begin{pmatrix}
-\sin \phi \sin{k} \\
\cos \phi \sin{k} \\
-\sin{(2 \theta)}  \cos{(k)}
\end{pmatrix},
\end{eqnarray}
and the damping spectra for the correlation functions are specified with ($\kappa =1$; cf. Eq. (\ref{spectras}))
\begin{eqnarray}
\kappa^{(c)}_k &=& 1+ \cos{(2 \theta) \cos{k}}, \,\,|\vec m_k^{a (c)}| =0, \\\nonumber
\kappa^{(n)}_k &=& 1, \,\, |\vec m_k^{a (n)}| = | \cos(2\theta) \cos k| .
\end{eqnarray}
Both cases reduce to the ideal case Eq. (\ref{equ:jump}) for $\theta = \pi/4 + s \pi, \phi=0$ ($s$ integer), and feature damping gap closing points at $\theta =s\pi/2$. The steady state in both cases is current free. The nature of these gap closing points is conveniently discussed by studying the relations of the steady vector $\vec n_k $ in their vicinity. In particular, we find the transformation rules  (with parameterization $\theta = \pi/2 +\delta\theta$, and restricting to the interval $\pi/4 \leq \theta \leq 3\pi/4$ without loss of generality)
\begin{eqnarray}\label{VecTraf}
\hspace{-0.2cm} \vec{n}^{(c)}_{k}(-\delta \theta) = S_z \vec{n}^{(c)}_{k}(\delta \theta) , \, \vec{n}^{(n)}_{k}(-\delta \theta)= - S_z \vec{n}^{(n)}_{k}(\delta \theta) 
\end{eqnarray}
for arbitrary $\phi$ with $S_z =\mathrm{diag}  (-1,-1,1)$. 
Thus, the steady state vectors on both sides of the transition point relate by reflections of two of their components with the third one fixed for all modes $k$, leading to the conclusions drawn in the text. In addition, from the transformation properties alone we can deduce basic statements on thermodynamic properties of the system at the transition point. In the first case,  the situation is thermodynamically trivial because the system is either fully filled or empty: at the transition point $\delta\theta =0$, by Eq. (\ref{VecTraf}) the vector points in $\pm z$-direction for all $k$. The additional constraint of purity only allows for $\vec n_k = (0,0,\pm 1)$ for each $k$. A steplike change of the sign for some values of $k$ (Fermi surface) is not possible in our context, since such step function in momentum space could only be synthesized from $u_k,v_k$ which are highly nonlocal in position space, contradicting quasilocality of the Lindblad operators. Thus, $\vec n_k = \pm \vec e_z$ with constant sign for all $k$, corresponding to an average filling $\bar n_k = \tfrac{1}{2} (1 - n_{z,k}) =0$ or 1. The situation is different for the second case: At the gap closing point, the vector $\vec n_k$ points in the $\pm y$-direction for all $k$, and there is no constraint on the purity. In fact, the system is half filled, $\bar n_k = \int \tfrac{dk}{2\pi} \bar n_k = \tfrac{1}{2} $, allowing thermodynamic observables to be properly defined.

\end{document}